\documentclass[10pt,journal,compsoc]{IEEEtran}

\ifCLASSOPTIONcompsoc
\usepackage[nocompress]{cite}
\else
\usepackage{cite}
\fi

\usepackage{dsfont}
\usepackage{amssymb}
\usepackage{subfig}
\usepackage{graphicx, amssymb, amsmath}
\usepackage{extarrows}
\usepackage{algorithm} %format of the algorithm
\usepackage[noend]{algpseudocode}
\usepackage{multirow} %multirow for format of table
\usepackage{color}
\usepackage{amsmath}
\usepackage{url}
\emergencystretch=\maxdimen
\IEEEoverridecommandlockouts 
\begin{document}

\title{Cooperative Edge Caching in User-Centric Clustered Mobile Networks}

\author{Shan~Zhang,~\IEEEmembership{Member,~IEEE,}
	Peter~He,~\IEEEmembership{Member,~IEEE,}
	Katsuya~Suto,~\IEEEmembership{Member,~IEEE,}
	Peng~Yang,~\IEEEmembership{Student~Member,~IEEE,}
	Lian~Zhao,~\IEEEmembership{Senior~Member,~IEEE,}	
	and~Xuemin~(Sherman)~Shen,~\IEEEmembership{Fellow,~IEEE}% <-this % stops a space
	\IEEEcompsocitemizethanks{\IEEEcompsocthanksitem Shan~Zhang, Peter~He, Katsuya~Suto and Xuemin~(Sherman)~Shen are with the Department of Electrical and Computer Engineering, University of Waterloo, 200 University Avenue West, Waterloo, Ontario, Canada, N2L 3G1.\protect\\
		E-mail: \{s327zhan, z85he, ksuto, sshen\}@uwaterloo.ca
	\IEEEcompsocthanksitem Peng~Yang is with the School of Electronic Information and Communications, Huazhong University of Science and Technology, Wuhan, China.\protect\\
		E-mail: yangpeng@hust.edu.cn
	\IEEEcompsocthanksitem Lian~Zhao is with the Department of Electrical and Computer Engineering, Ryerson University, Ontario, Canada, M5B 2K3.\protect\\
		E-mail: l5zhao@ryerson.ca}% <-this % stops an unwanted space
	%\thanks{This work is sponsored in part by the National Basic Research Program of China (973 Program: 2012CB316001), the National Science Foundation of China (NSFC) under grant No. 61201191 and No. 61401250, the Creative Research Groups of NSFC under grant No. 61321061, and Hitachi R\&D Headquarter.}% <-this % stops a space
	\thanks{Part of this work has been accepted by IEEE GLOBECOM~2017 \cite{mine_GC17_cache}.}
}

%%%\maketitle
%\markboth{IEEE Transactions on Mobile Computing,~Vol.~~, No.~~, May~2017}%
%{Zhang \MakeLowercase{\textit{et al.}}: Cooperative Edge Caching in User-Centric Clustered Mobile Networks}

\IEEEtitleabstractindextext{%
\begin{abstract}
	
	With files proactively stored at base stations (BSs), mobile edge caching enables direct content delivery without remote file fetching, which can reduce the end-to-end delay while relieving backhaul pressure.
	To effectively utilize the limited cache size in practice, cooperative caching can be leveraged to exploit caching diversity, by allowing users served by multiple base stations under the emerging user-centric network architecture. 	
	This paper explores delay-optimal cooperative edge caching in large-scale user-centric mobile networks, where the content placement and cluster size are optimized based on the stochastic information of network topology, traffic distribution, channel quality, and file popularity.
	%The design challenge is caching diversity and spectrum efficiency depend on the cluster size and content placement.
	Specifically, a greedy content placement algorithm is proposed based on the optimal bandwidth allocation, which can achieve $(1-{1/e})$-optimality with linear computational complexity.
	In addition, the optimal user-centric cluster size is studied, and a condition constraining the maximal cluster size is presented in explicit form, which reflects the tradeoff between caching diversity and spectrum efficiency.
	Extensive simulations are conducted for analysis validation and performance evaluation.
	Numerical results demonstrate that the proposed greedy content placement algorithm can reduce the average file transmission delay up to 45\% compared with the non-cooperative and hit-ratio-maximal schemes.
	Furthermore, the optimal clustering is also discussed considering the influences of different system parameters.
	%Furthermore, both analytical and numerical results reveals the tradeoff between caching diversity and spectrum efficiency in optimization of cooperative caching, considering the practical wireless transmission properties and radio resource constraints.
	
\end{abstract}

}

\maketitle
\IEEEdisplaynontitleabstractindextext

\IEEEpeerreviewmaketitle

\IEEEraisesectionheading{\section{Introduction}\label{sec:introduction}}

\IEEEPARstart{T}{o} accommodate the ever increasing mobile traffic demand, small cell base stations (SBSs) are expected to be ultra-densely deployed for extensive spatial spectrum reuse in the next generation (5G) networks and beyond \cite{ultra_dense_SC_2015}. %, 5G_Overview_JSAC14_JAndrews}.
As networks are further densified, deploying ideal backhaul for each SBS becomes impractical due to the high cost, leading to possible backhaul congestions and performance degradation \cite{Ge14_5G_backhaul_mag, Fang15_information_centric_survey}.
To relieve the backhaul pressure, mobile edge caching has been proposed to store contents at the edge of networks (e.g., SBSs or end devices) in addition to remote servers, whereby contents can be directly delivered through wireless transmission without backhaul or core network transmissions \cite{Poularakis17_cache_mobility_TMC, GreenDelivery_SZhou_2015, Bastug14_cache_framework_BS_D2D_mag, Yang16_catalyzing_cloud_fog_SDN, Wang16_D2D_caching_TVT}.
By exploiting the similarity of requested contents, mobile edge caching has the potential to reduce backhaul capacity requirement to 35\% \cite{traffic_relieve_caching}. %\cite{Wang14_cache_framework_wireless_mag}
In practical systems, the performance of mobile edge caching can be constrained by the limited cache size \cite{Liu16_EE_cache_JSAC, bacstuug2016delay}.
A straightforward solution is to design more effective content placement schemes, by exploring the information of content popularity and user preferences \cite{Wang16_cache_mobility_mag}.
Furthermore, cooperative caching can be leveraged to enlarge the set of cached files, which enables users served by multiple SBSs to exploit caching diversity in space \cite{Yin06_cooperative_caching_early, Chow07_cooperative_caching_early}.
{In fact, mobile networks are now evolving from the conventional cellular topology to a de-cellular user-centric structure, where each user can be served by a dynamically formed cluster of SBSs for quality of experience (QoE) enhancement \mbox{\cite{Chen16_UUDN_mag, Bao16_handoff_rate_user_centric_infocom, Song12_multi_radio_TWC, Nie16_EE_clustering_user_centric_HetNet_JSAC}}.
In this case, different SBSs can store diverse contents in a cooperative manner, such that users have higher probability to obtain the requested contents locally from the caches of clustered SBSs \mbox{\cite{Song15_single_UE_cluser_cache_BER_ICC}}.
However, cooperative caching may sacrifice spectrum efficiency due to the enlarged transmission distance.
Specifically, increasing the cluster size improves the content hit ratio, but users are served by farther SBSs with higher path loss, degrading spectrum efficiency.
The tradeoff relationship between caching diversity and spectrum efficiency in terms of the cluster size can bring significant challenges for edge caching, which has not been well studied in literature \mbox{\cite{Chen17_CoMP_cluster_cache_TWC}}.}
%Cooperative caching requires users to be served by multiple candidate SBSs, which meets the trend of mobile networks \cite{Ismail13_multihoming_twc, Hong13_BS_clustering_JSAC}.

This paper investigates delay-optimal cooperative edge caching in large-scale user-centric clustered mobile networks considering the constrained cache size and radio resources, based on the stochastic information of network topology, traffic distribution, channel quality, and file popularity.
Specifically, we focus on two fundamental problems: 1) content placement, and 2) SBS clustering.
Coded caching is adopted, whereby each user can fetch coded segments for decoding from the caches of the candidate SBSs in cluster.
For the content placement design, an optimization problem is formulated to determine the ratio of cached segments for different files, aiming at minimizing the average file transmission delay.
In addition, the bandwidth allocation is jointly optimized for load balancing, since different content placement results in different traffic distributions.
The formulated problem is challenging for following reasons.
Firstly, the accurate form of average file transmission delay cannot be derived due to the multi-dimensional randomness of network topology, traffic distribution, and channel quality.
Secondly, the problem is a NP-hard mixed integer programming problem, as content placement and bandwidth allocation determines the average file transmission rate in a coupled manner. 
Thirdly, the average file transmission rate has a piecewise structure with respect to content placement, introducing additional complexity for delay performance analysis.
{{By applying the theory of stochastic geometry, we obtain the average transmission rate with conservative approximation, whereby two subproblems need to be addressed, i.e., bandwidth allocation and content placement.
The bandwidth allocation subproblem is convex and thus is solved by employing Lagrange multiplier method.
To address the non-convex content placement subproblem, we prove the piecewise objective function has monotone submodular property, and thus propose a greedy algorithm which can achieve $(1-\frac{1}{e})$-optimality with linear complexity.}}
For the SBS clustering, an explicit condition constraining the maximal cluster size is obtained, which reveals the tradeoff between content diversity and spectrum efficiency with respect to system parameters (including SBS density, backhaul delay and content popularity). 
{{Extensive simulations are conducted to validate the analysis as well as evaluate the performance of proposed algorithm, based on both real-world YouTube data trace and the classic Zipf content popularity distribution.
Two typical caching schemes, non-cooperative and hit-ratio-maximal, are adopted to demonstrate the tradeoff between content diversity and spectrum efficiency in user-centric clustered caching. 
Simulation results show that the non-cooperative scheme guarantees highest spectrum efficiency with lowest content ratio, which performs well in case of low network density or large cache size; whereas the hit-ratio-maximal caching scheme provides maximal hit rate but sacrifices the spectrum efficiency, which is more advantageous in dense networks with congested backhaul and small cache size. 
The proposed greedy caching scheme always outperforms the other two schemes, by adjusting content placement and cluster size to balance content diversity and spectrum efficiency.
Furthermore, the influences of important system parameters are also studied, including cache size, SBS density, backhaul delay.}}

The main contributions of this work are as follows.
\begin{enumerate}
	\item {The content placement, SBS clustering, and bandwidth allocation have been jointly optimized in large-scale user-centric mobile networks considering the radio resource constraints, based on the stochastic information of network topology, traffic distribution, channel quality, and file popularity.}
	\item {The average file transmission delay is derived with conservative approximation, and proved to have monotone submodular property with respect to content placement. Accordingly, a low-complexity greedy content placement algorithm is proposed with $(1-{1/e})$-optimality performance guarantee.}
	\item {The bandwidth allocation is optimized to minimize average file transmission delay, which can match radio resource to the traffic load distribution resulting from content placement.}
	\item {The delay-optimal SBS clustering for cooperative caching is obtained, and the influences of important system parameter are analyzed.}
	\item {Both analytical and numerical results reveal the underneath tradeoff relationship between content caching diversity and spectrum efficiency in cooperative caching, which can provide insightful guidelines to dynamic SBS clustering and content placement with the variations of traffic loads and operational scenarios.}
\end{enumerate}

 The remainder is organized as follows. Section~\ref{sec_review} reviews related work on cooperative caching, and Section~\ref{sec_system_model} introduces the system model and problem formulation. The average file transmission delay is analyzed in Section~\ref{sec_analysis}, based on which the optimal bandwidth allocation is derived. In Section~\ref{sec_submodular}, a greedy content placement algorithm is proposed. Finally, Section~\ref{sec_simulation} shows simulation results, and Section~\ref{sec_conclusion} concludes the paper.

%%%%%%%%%%%%%%%%%%%%%%%%%%%%%%%%%%%%%%%%%%%%%%%%%%%%%%%%%%%%%%%%%%%%%%%%%%%%%%%%%%%%%%%%%%%%%%%%%%%
\section{Literature Review}
	\label{sec_review}
	Although the content placement problem has been extensively studied in wired networks, the design of mobile edge caching is relatively underdevelopped, due to the features of user mobility, link connectivity and channel quality \cite{Yin06_cooperative_caching_early, He16_video_routing_cache}.
	When each BS provides service independently, the popularity-based cache placement scheme (i.e., each BS stores the most popular contents) has been widely adopted to maximize the content hit ratio \cite{Gong17_push_TWC}.
	The cooperation among BSs can further enhance caching efficiency, which is also more challenging since the caching decision of one BS can be influenced by neighboring BSs \cite{Li16_cache_min_weighted_load_ICC}.
	Existing works on cooperative mobile edge caching can be classified into two categories based on the utilized system information, i.e., complete priori information and stochastic information.
	
	With complete priori information of user connectivity and channel quality, both centralized and distributed cooperative content placement schemes have been designed \cite{Shanmugam13_content_place_TIT,  Li15_distributed_caching, Tran16_octopus, Jiang17_SBS_UE_caching_TMC}.
	Shanmugam \emph{et al.} have investigated the centralized cooperative content placement problem to minimize the average file downloading delay in a network consisting of one MBS and several cache-enabled SBSs, considering both coded and uncoded caching \cite{Shanmugam13_content_place_TIT}.
	%The problem has been proved to be NP-hard for the uncoded caching case, and a greedy algorithm is proposed within a factor 2 of optimum.
	%If coded caching is enabled, the problem becomes convex and can be reduced to linear programing.
	Li \emph{et al.} have proposed a distributed belief propagation algorithm for content placement, by applying the factor graph to describe network topology \cite{Li15_distributed_caching}.
	{{Tran \emph{et al.} have further combined the edge-based BS caching with core-based cloud caching in the Cloud-based Radio Access Networks (C-RANs), and designed a hierarchical content placement scheme to reduce file fetching delay and backhaul pressure {\cite{Tran16_octopus}}.
			The cooperative caching among BSs and users has been studied in {\cite{Jiang17_SBS_UE_caching_TMC}}, which can be formulated as an integer-linear programming problem and solved by the hierarchical primal-dual decomposition method.
			In addition to popularity-based caching, other works have also exploited the user mobility information for proactive content fetching and cache update, to provide seamless handover experience \mbox{\cite{EdgeBuffer15_mobility_caching_wowmom}}.
			Furthermore, caching schemes jointly consider content popularity and user mobility have been also designed \mbox{\cite{Vasilakos16_mobility_popularity_caching}}.
			By utilizing the complete information of network status, these algorithms can significantly improve network performance in terms of transmission delay, content hit rate, backhaul load, and user QoE.
			However, the limitation is that the cached contents need to be updated frequently with the variations of user requests or channel conditions, which may introduce considerable overhead and backhaul loads in practice.
			Furthermore, these algorithms are usually designed on a case-by-case basis for small-scale networks, which cannot provide general design guidelines for practical networks.}}
	
	Cooperative caching has also been studied based on stochastic network information.
	By modeling the cache-enabled BSs as a homogeneous Poisson Point Process (PPP), the caching probabilities of different files have been optimized to maximize the content delivery success probability, where each user is served by the nearest BS with requested contents in cache \cite{chae16_random_cache_PPP_TWC}.
	Our previous work has shown the tradeoff between content diversity and spectrum efficiency in cooperative caching, when the user can be steered to the second nearest SBSs (i.e., user-centric cluster size set to 2) \cite{mine_GC17_cache}.
	For multi-tier heterogeneous networks, an iterative heuristic content placement algorithm has been proposed, where each network tier makes decisions based on the contents cached at other tiers \cite{Serbetci17_hit_rate_multi_PPP_WCNC}.
	Furthermore, the long-term cache instance deployment problem has been studied to determine the optimal cache size of SBSs and MBSs for the given storage resource budget, where the SBS-tier cache the most popular contents and the MBS-tier cache the less popular ones for load balance \cite{mine_hierarchical_cache_TVT}.
	Very recent works have explored the cooperative caching in user-centric clustered networks  \cite{Song15_single_UE_cluser_cache_BER_ICC, Chen17_CoMP_cluster_cache_TWC}, where insightful analytical results have been obtained on the tradeoff between caching diversity and channel diversity.
	However, \cite{Song15_single_UE_cluser_cache_BER_ICC, Chen17_CoMP_cluster_cache_TWC} only considered the single user case and ignored the constraints of radio resources.
	In fact, radio resource management has great impact on the performance of edge caching, which should be jointly optimized to further enhance system performance \cite{Wang16_UA_CP_Access, Khreishah16_cache_RA_JSAC, Ismail13_multihoming_twc}. %\cite{Cui16_CA_RA_HetNet_GC} 
	
	{{This paper investigates the joint optimization of content placement, SBS clustering, and bandwidth allocation in large-scale user-centric mobile networks, based on the stochastic network information.
			The novelty is three-fold.
			Firstly, by jointly optimizing bandwidth allocation, we conduct network-level cooperative caching design with the constraint of radio resources taken into consideration, which has been ignored in the existing clustered cooperative caching studies focusing on user-aspect performance {\cite{Song15_single_UE_cluser_cache_BER_ICC, Chen17_CoMP_cluster_cache_TWC}}.
			Secondly, through the design of content placement and cluster size, we reveal the underneath tradeoff relationship between content diversity and spectrum efficiency in an analytical way, which has not been well investigated in existing literature.
			Thirdly, both the analytical and numerical results demonstrate that the cluster size should be adjusted based on the traffic load and network status to balance the content diversity and spectrum efficiency, whereas most existing works have adopted the constant cluster size only based on the received signal strength.
		}}
%%%%%%%%%%%%%%%%%%%%%%%%%%%%%%%%%%%%%%%%%%%%%%%%%%%%%%%%%%%%%%%%%%%%%%%%%%%%%%%%%%%%%%%%%%%%%%%%%%%
\section{System Model and Formulation}
    \label{sec_system_model}
    \begin{figure}[!t]
    	\centering
    	\includegraphics[width=2.5in]{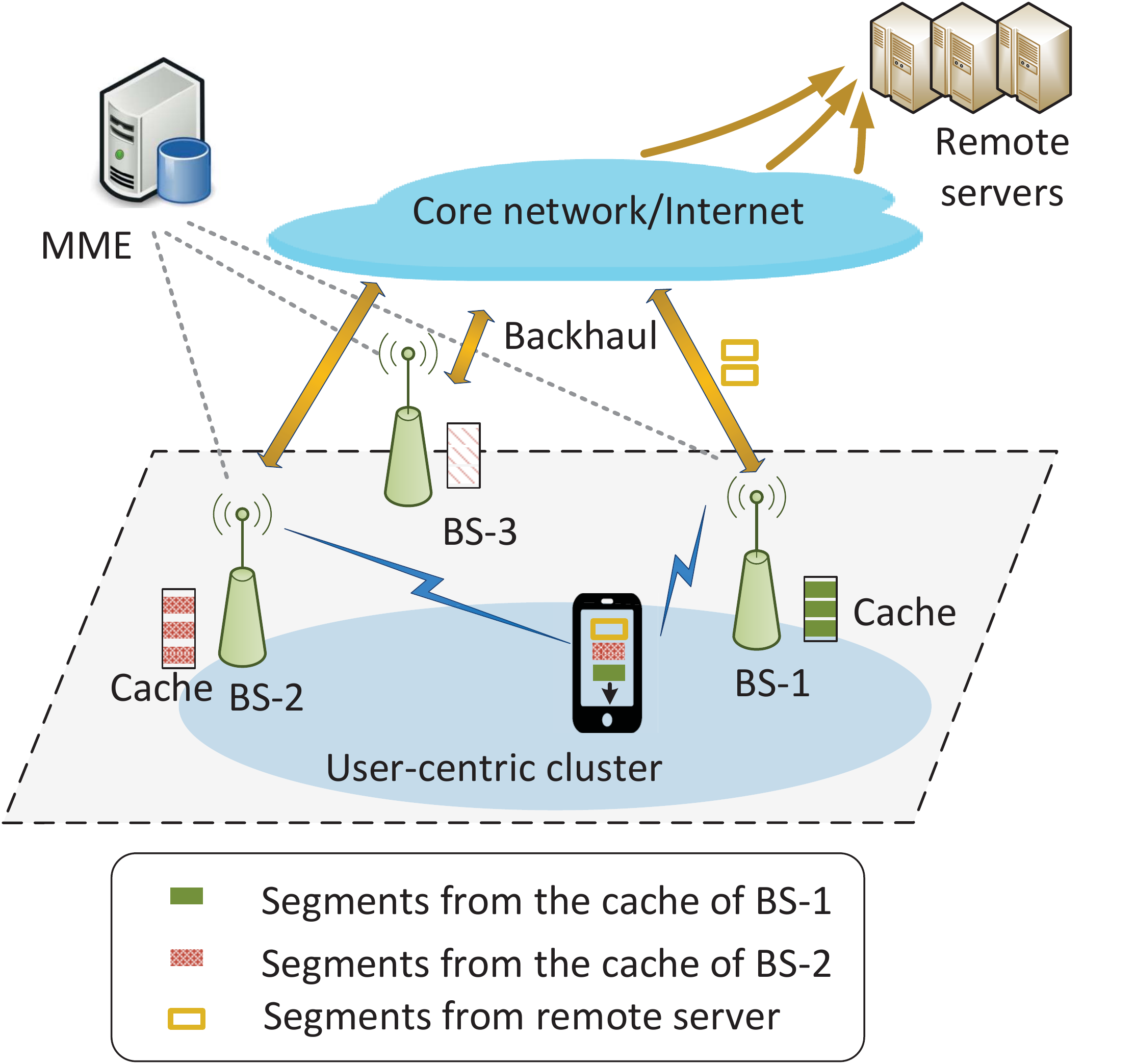}
    	\caption{{{Mobile edge caching with user-centric clustered services.}}}
    	\label{fig_scenario}
    \end{figure}

    We consider a {{homogeneous}} mobile network with edge caching, where contents can be partially or completely stored at each SBS after being coded into segments, as shown in Fig.~\ref{fig_scenario}.
    {{The contents are characterized by popularity based on the hit rate, corresponding to diverse mobile services such as videos streaming, HD map, social media, news, software update. }}
    When a user raises a content request, it can be served by a cluster of candidate SBSs, depending on the content caching states.
    If the requested content is cached, the user fetches the coded segments directly from candidate SBSs in ascending order of transmission distance, until the obtained segments are sufficient for file decoding.
    In addition, if the obtained segments from the caches of all candidate SBSs are insufficient for decoding, the nearest SBS will fetch the remaining ones from remote servers through backhaul, and then deliver to the user through wireless transmission.
    If the requested content is not stored in cache, the nearest SBS will fetch the whole content from remote servers. 
    Therefore, the transmission delay depends on the content placement.
    %This section introduces the system model to prepare the analysis of file transmission delay.
    The main notations are listed in Table~\ref{tab_notation}.
    
    \subsection{Cooperative Content Caching}
    
    {{The distributions of SBSs and end users are modeled as independent PPPs of densities $\rho$ and $\lambda$, respectively, for tractable analysis \mbox{\cite{JAndrews2011tractable}}.}}
    Denote by $\mathcal{F}=\left\{ 1,2,\cdots, f,\cdots, F \right\}$ the file library, where $F$ is the total number of files.
    Denote by $\mathcal{Q} = \left\{ q_1, q_2,\cdots,q_f,\cdots,q_F \right\}$ the file popularity distribution, where $q_f>0$ is the probability that the requested content is file-$f$ ($\sum_{f=1}^{F} q_f = 1$).
    We exploit rateless fountain coding for cooperative caching \cite{Mackay05_fountain}. 
    Each file is encoded by fountain coding into independent segments, which are equally deployed to all SBSs, i.e., each SBS holds $c_f$ encoded segments of file-$f$\footnote{{{The identical file caching at SBSs is adopted since the SBSs are homogeneous with the same system parameters (transmit power, bandwidth, cache size and backhaul capacity).}}}.
    Users can decode file-$f$ by collecting $s_f$ encoded segments from SBSs.
    The coded segments are considered to have the same size of $L$ in bits, and thus $s_f L$ reflects the size of file-$f$ in bits.
    Due to the limited cache size, each SBS can store $C$ coded segments at most, i.e., $\sum_{f=1}^{F} c_f \leq C$.
    
    \begin{table}[!t]
    	\caption{Notation table}
    	\label{tab_notation}
    	\centering
    	\begin{tabular}{cc}
    		\hline
    		\hline
    		Notation & Definition/Description  \\
    		\hline
    		$\rho$ & Density of SBSs \\
    		$\lambda$ & Density of active users\\
    		$F$ & The number of files in the file library\\
    		$f$ & File index, $f=1,\cdots,F$\\	
    		$q_f$ & File popularity of file-$f$\\
    		$L$ & Length of each coded \\
    		$s_f$ & The number of coded segments of file-$f$\\
    		$c_f$ & The number of coded segments of file-$f$ cached\\
    		$C$ & Cache size of each SBS\\
    		$K$ & SBS cluster size \\
    		$B_k$ & The $k$th nearest SBS, $k=1,2,\cdots,K$ \\
    		$P_{k,f}$ & Downloading ratio at $B_k$ if file-$f$ is requested\\
    		$\Omega_k$ & Average ratio of traffic served by $B_k$ \\
    		$W$ & Available system bandwidth\\
    		$\varphi_k$ & Ratio of bandwidth allocated to Group-$k$ users\\
    		$N_k$ & Number of Group-$k$ users served by a SBS\\
    		$\zeta$ & Spectrum efficiency of Group-$k$ users\\
    		$R_k$ & Transmission rate of Group-$k$ users \\
    		$P_\mathrm{T}$ & Transmit power of SBSs\\
    		$d_k$ & Transmission distance of Group-$k$ users\\
    		$\alpha$ & Path loss coefficient\\
    		$\sigma^2$ & Addictive Gaussian noise power density\\
    		$I_k$ & Inter-cell interference received by Group-$k$ users\\
    		$\bar{D}$ & Average transmission delay \\
    		\hline
    		\hline
    	\end{tabular}
    \end{table}
    
    \begin{figure}[!t]
    	\centering
    	\includegraphics[width=1.7in]{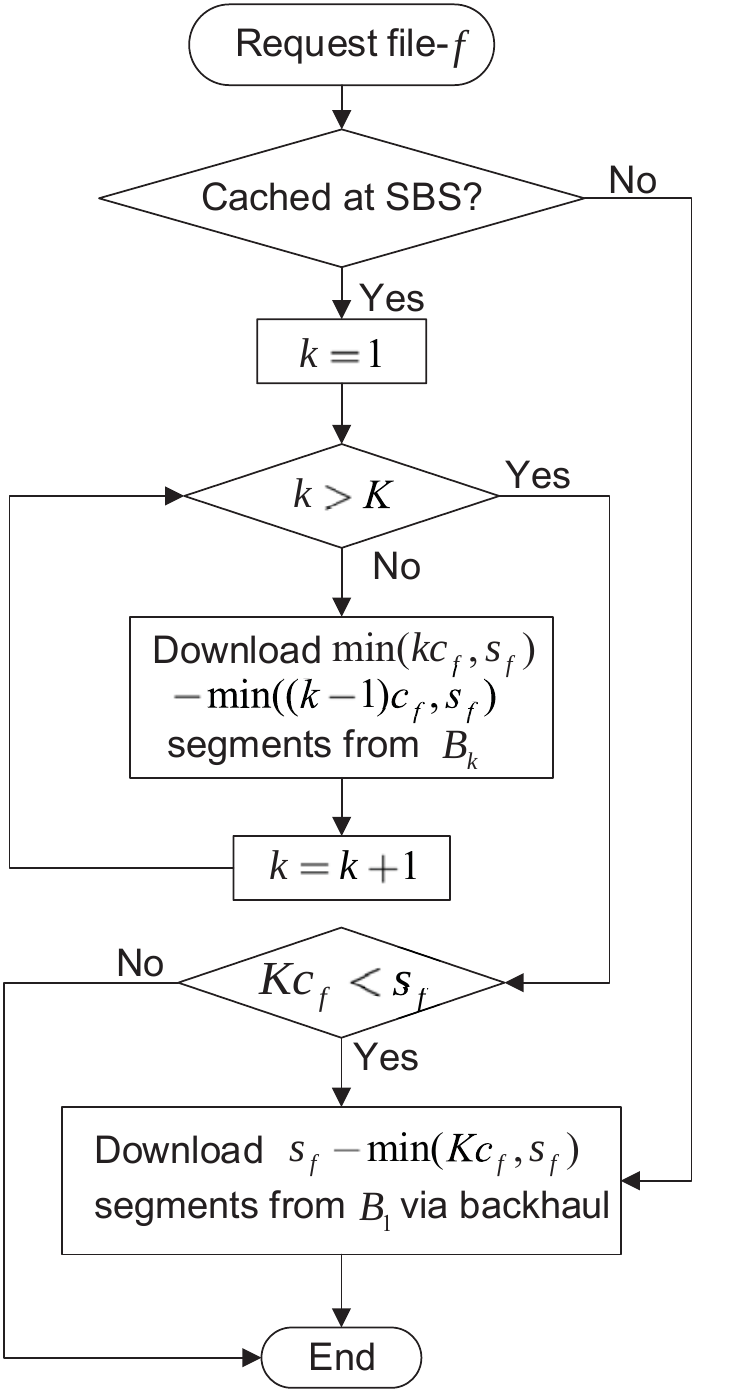}
    	\caption{Service process with user-centric clusters.}
    	\label{fig_service_flow}
    \end{figure}

    \begin{figure*}[!t]
    	\centering
    	\includegraphics[width=6in]{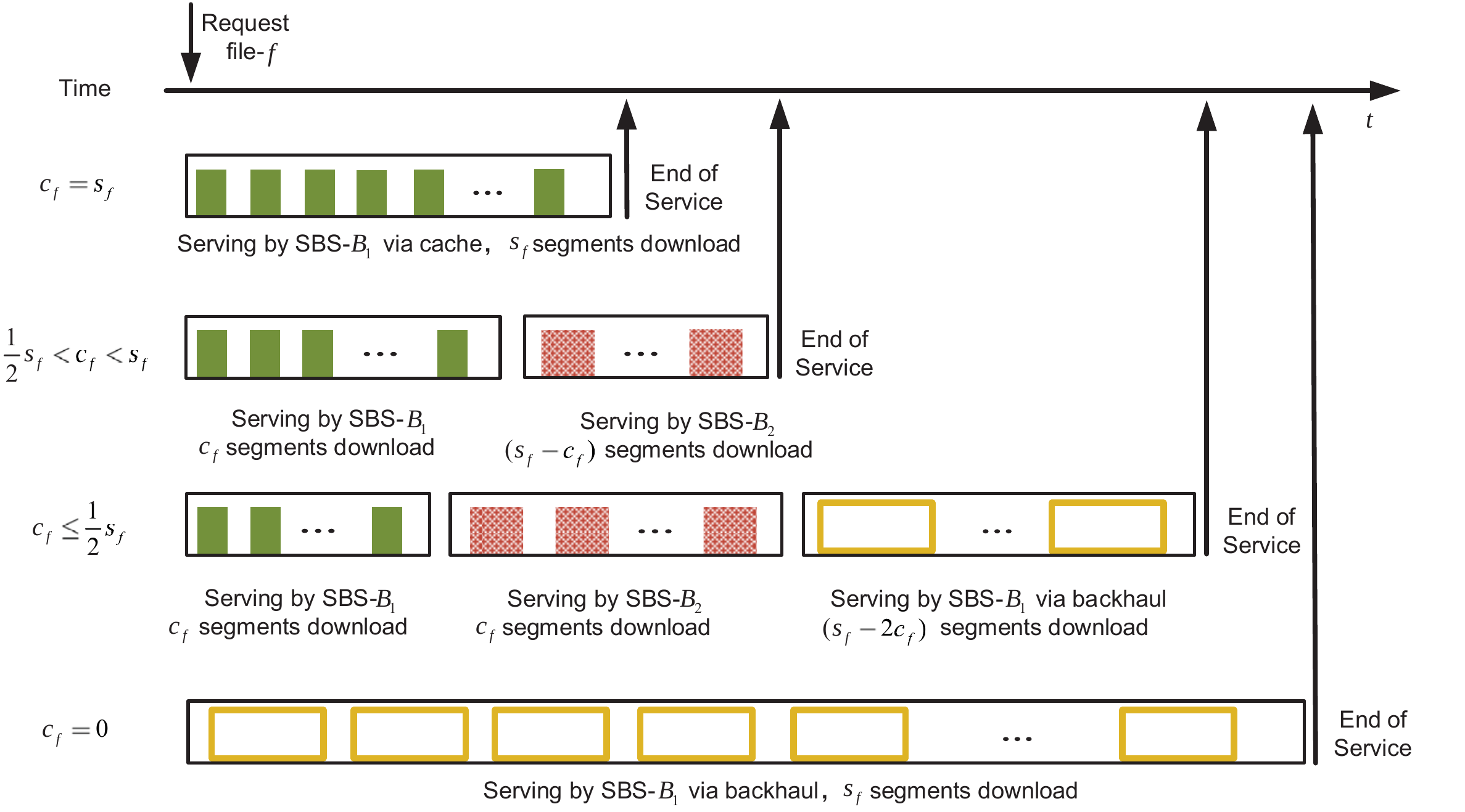}
    	\caption{{Segment downloading illustration when cluster size is 2.}}
    	\label{fig_service_process_horizontal}
    \end{figure*}
    
    {{Consider a typical user $u$, served by $K$ SBSs which have the shortest transmission distance.
    		As the SBSs are homogeneous with the same system parameters like transmit power, the distance-based SBS clustering can provide highest transmission rate on average.
    		In practical systems, the transmission distance can be estimated by the user-side channel measurement, and the information is reported to the Mobility Management Entity (MME) which makes decisions on user association and service.}}
    Denote by $\mathcal{B}_u = \{B_1,B_2,\cdots,B_k,\cdots,B_K\}$ the set of candidate SBSs, which are sorted according to transmission distance in ascending order without loss of generality.
    {{Suppose user-$u$ requests file-$f$, and it will fetch the cached segments from SBS-$B_1$ to SBS-$B_K$ successively, until the number of obtained segments reaches $s_f$.
    		If $K c_f < s_f$, SBS-$B_1$ will fetch the remaining $\left( s_f - K c_f \right)$ segments from remote servers via backhaul, which does not break cell load balancing.
    		In addition, such service scheme can minimize the average transmission delay, without the instant information of channel status or cell load.}}
    User-$u$ can get $\left[ \min\left( k c_f, s_f \right) - \min\left( (k-1) c_f, s_f \right) \right]$ segments from SBS-$B_k$, and SBS-$1$ needs to fetch $\left[ s_f - \min\left( K c_f, s_f \right) \right]$ segments from remote servers.
    The service flow chart is given as Fig.~\ref{fig_service_flow}.
    When the cluster size is $K=2$, Fig.~\ref{fig_service_process_horizontal} illustrates the detailed service process with different cache ratio, corresponding to Fig.~\ref{fig_scenario}.
    Notice that the average downloading delay is lower at SBS-$B_1$ compared with that at SBS-$B_2$, which will be analyzed in details later.
    In addition, the delay will further increase when fetching the remaining segments from remote servers, due to the backhaul transmission.
    Thus, increasing $c_f$ helps to reduce average transmission delay of file-$f$, as shown in Fig.~\ref{fig_service_process_horizontal}.
    As a results, the content placement should be well designed to minimize delay, under the constraint of cache size $\sum_{f=1}^{F} c_f \leq C$. 
    
    Denote by $P_{k,f}$ the ratio of segments that user-$u$ can get from the cache of SBS-$B_k$, given by
    \begin{equation}
    \label{eq_hit_ratio_f}
    P_{k,f} = \frac{1}{s_f} \left[\min\left( k c_f, s_f \right) - \min\left( (k-1) c_f, s_f \right) \right].
    \end{equation}
    For notation simplicity, denote by $P_{K+1,f}$ the ratio of remaining contents fetched from remote servers:
    \begin{equation}
    \label{eq_hit_ratio_miss}
    P_{K+1,f}  \triangleq 1 - \sum_{k=1}^{K} P_{k,f} =  1 - \min\left( K c_f, s_f \right) \frac{1}{s_f}.
    \end{equation}
    Therefore, the average content hit ratio at the $k$th candidate SBS is given by
    \begin{equation}
    \label{eq_group_ratio}
    \Omega_k = \sum\limits_{f=1}\limits^{F} q_f P_{k,f},
    \end{equation}
    where $k=1,2,\cdots,K$.
    Denote by $\Omega_{K+1}$ the average content miss ratio: $\Omega_{K+1} = 1 - \sum_{k=1}^{K} \Omega_k$.
    %Notice that $\Omega_k$ represent the average ratio of content a user can get from the $k$th candidate SBS.
    
    \begin{table}[!t]
    	\caption{Example of user-centric cluster}
    	\label{tab_traffic_classify}
    	\centering
    	\begin{tabular}{|c||c|c|c|c|c|c|}
    		\hline
    		& user-1 & user-2 & user-3 & user-4 & user-5 \\ 
    		\hline
    		SBS-1 & $B_1$ & $B_2$ & $B_1$ & $\times$ & $\times$  \\ 
    		SBS-2 & $B_2$ & $\times$ & $\times$ & $B_1$ & $B_2$  \\
    		SBS-3 & $\times$ & $B_1$ & $B_2$ & $B_2$ & $B_1$ \\ 
    		\hline
    	\end{tabular}
    \end{table}

    For a typical SBS, it can hold different rankings to the served users, varying from the 1st to the $(K+1)$th.
    Thus, from the perspective of SBSs, the served users can be classified into $(K+1)$ groups, based on the ranking of the serving SBSs.
    {{From the perspective of an individual user, it can belong to different groups at different time, since it fetches segments from different SBSs successively.
    		User-$u$ belongs to Group-$k$ when it is fetching file from SBS $B_k$, i.e., the $k$th candidate SBS.}}
    Table~\ref{tab_traffic_classify} gives an example of network topology when there are 3 SBSs serving 5 users with cluster size of 2, where $B_k$ means the SBS is the $k$th nearest SBS for the corresponding user ($\times$ means the SBS is not connected).
    For illustration, SBS-2 is the second nearest SBS for user-1 and user-5, and is the nearest SBS for users-4.
    As a result, SBS-2 can serve three users, user-4 belongs to the first user group when downloading cached segments, user-1 and user-5 belong to the second user group, while user-4 belongs to the third group when downloading remaining segments via the backhaul.
    
    {{The main difference between the groups is the wireless transmission rate.
    		Specifically, the transmission distance increases with the group index $k$, leading to higher path loss and lower transmission rate on average.
    		According to Eq.~(\ref{eq_group_ratio}), $\Omega_k$ also denotes the probability that a user is being served by the $k$th candidate SBS, i.e., belonging to Group-$k$.
    		As the user distribution follows a PPP of density $\lambda$, the distribution of Group-$k$ users also follows a PPP of density $\Omega_k \lambda$.
    		Define $\{\Omega_k\}$ as the \emph{group load distribution}, to denote the traffic load distribution among different groups, which is a key factor influencing the average file transmission delay.}}

    \subsection{Wireless Transmission}
    
    The transmission rate should be guaranteed for successful file delivery.
    When user-$u$ is served by SBS-$B_k$, the file transmission rate is given by
    \begin{equation}
    R_k = w_{u,k} \log_2 \left(1 + \zeta_k\right),
    \end{equation}
    where $w_{u,k}$ is the allocated bandwidth to user-$u$ and $\zeta_k$ is the received signal to interference and noise ratio (SINR)\footnote{$\zeta_{K+1}=\zeta_1$.}:
    \begin{equation}
    \zeta_k = \frac{P_\mathrm{T} {d_k}^{-\alpha}}{\sigma^2 + I_k},
    \end{equation}
    $P_\mathrm{T}$ is the SBS transmit power density, $d_k$ is the distance from user-$u$ to SBS-$B_k$, $\alpha$ is the path loss exponent, $\sigma^2$ is the additive Gaussian noise power density, and $I_k$ denotes the inter-cell interference when served by SBS-$B_k$.
    {{With the advanced interference mitigation techniques (like spectrum reuse, power control, and interference alignment), the inter-cell interference can be treated as constant noise \mbox{\cite{Intf_noise_half}}. Then, $I_k$ can be interpreted as the efficiency of interference mitigation. For example, $I_k=0$ if the interference is canceled completely, which usually requires perfect channel information. Larger $I_k$ means less effective interference mitigation, which can happen with incomplete channel information, lower received signal strength, extensive spatial spectrum use \mbox{\cite{Intf_noise_gap}}.}}
    
    Denote by $W$ the total available system bandwidth, and $\phi_k$ the ratio of bandwidth allocated to Group-$k$ users, where $\phi_k \geq 0$ and $\sum_{k=1}^{K+1} \phi_k \leq 1$.
    When user-$u$ is being served by SBS $B_k$, it belongs to Group-$k$ and the corresponding transmission rate is given by
    \begin{equation}
    \label{eq_r_k_band_allocation}
    R_k = \frac{\phi_k W}{N_k} \log_2 \left(1 + \zeta_k\right),
    \end{equation}
    where $N_k$ is the number of Group-$k$ users being served by SBS-$B_k$.
    Due to the random network topology and user location, $N_k$ and $\zeta_k$ are random variables.
    Thus, the average file transmission delay is given by
    \begin{equation}
    \label{eq_delay_define}
    \bar{D} = \sum_{k=1}^{K} \frac{ \Omega_k \bar{S} L }{\underset{\{N_k, d_k\}}{\mathds{E}} \left[ R_k \right]} + \left(\frac{\bar{S} L}{\underset{\{N_{K+1}, d_{K+1}\}}{\mathds{E}}\left[ R_k \right]}+\bar{D}_\mathrm{BH}\right) \Omega_{K+1},
    \end{equation}
    where $\underset{\{N_k, d_k\}}{\mathds{E}} \left[ R_k \right]$ denotes the expected transmission rate of Group-$k$ users in respect to the random variables $N_k$ and $d_k$, $\bar{D}_\mathrm{BH}$ is the average delay of backhaul file fetching, and $\bar{S}$ is the average length of requested files, i.e., $\bar{S}=\sum_{f=1}^{F} q_f s_f$.
    The probability distribution of $N_k$ depends on the group load distribution $\{\Omega_k\}$, which varies with content placement $\{c_f\}$ according to Eq.~(\ref{eq_group_ratio}).
    Accordingly, the bandwidth allocation $\{\phi_k\}$ should be adjusted to $\Omega_k$ for load balancing and effective resource utilization, based on the content placement.
    Therefore, we jointly optimize the content placement and bandwidth allocation, to minimize the average file transmission delay.

    \subsection{Problem Formulation}
    
    The delay-optimal cooperative caching problem can be formulated as follows:
    \begin{subequations}
    	\label{eq_P1}
    	\begin{align}
    	\underset{\{c_f\},\{\phi_k\}}{\min} ~~& \bar{D}, \\
    	\mbox{(P1)}~~~~\mbox{s.t.}~~~~~& \sum_{f=1}^{F} c_f  \leq C,\\
    	& \sum_{k=1}^{K+1} \phi_k \leq 1,\\
    	& c_f\in\{0,1,\cdots, s_f\}, ~~\forall f\in\mathcal{F},\\
    	& \phi_k \geq 0,~~k=1,2,\cdots,K+1,
    	\end{align}
    \end{subequations}
    where (\ref{eq_P1}b) reflects the limitation of cache capacity and (\ref{eq_P1}c) is due to the constrained system bandwidth.
    The optimization of content placement $\{c_f\}$ should consider the tradeoff between content diversity and spectrum efficiency.
    Increasing content diversity can improve content hit ratio, and the average backhaul transmission delay can be reduced as more users can get requested contents directly from the candidate SBSs.
    However, the spectrum efficiency degrades due to the increased transmission distance, introducing higher delay of wireless transmission.
    The optimization of bandwidth allocation (i.e., $\{\phi_k\}$) is to balance the resource and traffic demand among different groups, since different content placement leads to different group load distributions (i.e., $\{N_k\}$).
    
    Problem (P1) has two-fold challenges.
    Firstly, the average delay cannot be derived in closed form, due to the multi-dimensional randomness of network topology, traffic distribution, as well as file popularity.
    Accordingly, the relationship between average delay and content placement cannot be obtained directly.
    Secondly, as $\{c_f\}$ takes discrete values while $\{\phi_k\}$ is continuous, (P1) is a mixed integer programing problem and cannot be solved in polynomial time.
    To deal with the first challenge, we apply the stochastic geometry and analyze the average delay approximately based on the high SINR requirements of practical networks.
    As for the second challenge, (P1) can be solved in two steps: 1) \emph{Bandwidth Allocation Subproblem}: deriving the optimal bandwidth allocation for any given content placement; and 2) \emph{Content Placement with Optimal Bandwidth Allocation}: optimizing the caching problem by substituting the obtained optimal bandwidth allocation into (P1).
    In the following sections, the analysis of average file transmission delay, bandwidth allocation and content placement subproblems will be presented in details. 
%%%%%%%%%%%%%%%%%%%%%%%%%%%%%%%%%%%%%%%%%%%%%%%%%%%%%%%%%%%%%%%%%%%%%%%%%%%%%%%%%%%%%%%%%%%%%%%%%%%
\section{Average Delay Analysis}
    \label{sec_analysis}
    In this section, we analyze the average file transmission delay via stochastic geometry, based on which the bandwidth allocation subproblem is solved by applying the Lagrange multiplier method.
    
    \subsection{File Transmission Delay}
    
    {{Due to the multi-dimensional randomness of cell load, transmission distance and coverage area, the accurate transmission rate cannot be derived, posing challenges to the design of cooperative caching.
    		However, a lower bound of average file transmission rate can be approximately obtained in closed form, based on the theory of stochastic geometry, given as Lemma~1.}}
    
    \textbf{Lemma 1.} The average file transmission rate for Group-$k$ users has a lower bound, 
    \begin{equation}
    \label{eq_r_aver_k}
    \mathds{E}[R_k] \geq \frac{\phi_k W \rho}{\lambda \Omega_k } \left[ \log_2 \frac{P_\mathrm{T} (\pi \rho)^{\frac{\alpha}{2}}}{\sigma^2 + I_k} + \frac{\alpha}{2\ln 2} (\gamma - \sum\limits_{m=1}\limits^{k-1} \frac{1}{m})  \right],
    \end{equation}
    where $\gamma \approx 0.577$ denotes the Euler-Mascheroni constant.
    In addition, the equality holds when $\frac{\sigma^2 + I_k}{P_\mathrm{T}} \rightarrow 0$.
    
    %\emph{Proof.} Please refer to Appendix~\ref{appendix_lemma_2}.
    \emph{Proof.} {{As the received SINR and cell load can be considered as independent variables, the average file transmission rate can be approximated as:}}
    \begin{equation}
    \label{eq_appendix_A_1}
    \mathds{E}[R_k] = \frac{\phi_k W}{\ln 2 } \frac{\mathds{E} \left[\ln \left( 1 + \zeta_k \right)\right]}{\mathds{E}[N_k] }, %= \frac{\phi_k W \rho}{\lambda \Omega_k \ln 2} \mathds{E} \left[\ln \left( 1 + \zeta_k \right)\right],
    \end{equation}
    {{where $\mathds{E}[N_k] = \lambda \Omega_k/\rho$ since both users and SBSs follow PPPs.}}
    Furthermore, 
    \begin{equation}
    \label{eq_appendix_A_2}
    \mathds{E} \left[\ln \left( 1 + \zeta_k \right)\right] \geq -\alpha \mathds{E} \left[\ln \left(  d_k \right)\right] +\ln\left(\frac{P_\mathrm{T}}{I_k + \sigma^2}\right),
    \end{equation}
    where the equality holds if $\frac{I_k + \sigma^2}{P_\mathrm{T}}\rightarrow 0$ (i.e., high SINR).
    {{The cumulative distribution function (CDF) of the transmission distance $d_k$ is given by}}
    \begin{equation}
    \mathds{P}\left\{d_k \leq D \right\} = 1 - \sum_{m=0}^{k-1} \frac{(\pi D^2 \rho)^m}{m!}e^{-\pi D^2 \rho},
    \end{equation}
    i.e., the probability that there are at least $k$ SBSs within the circular of radius $d_k$ centered at a user.
    Thus,
    \begin{equation}
    \small
    \label{eq_appendix_lemma_2_1}
    \begin{split}
    & \mathds{E} \left[\ln \left(  d_k \right)\right] = \int_{0}^{\infty} \ln D \mbox{~d} (\mathds{P}\left\{d_k\leq D\right\})\\
    & = \int_{0}^{\infty}  2 \pi \rho D e^{-\pi \rho D^2}  \ln D \mbox{~d} D \\
    & ~~~~ - \int_{0}^{\infty} \ln D \mbox{~d} \left( \sum_{m=1}^{k-1} \frac{(\pi D^2 \rho)^m}{m!}e^{-\pi D^2 \rho} \right) \\
    & = \int_{0}^{\infty} \frac{1}{2} e^{-z} \ln z d z -  \int_{0}^{\infty} \frac{1}{2} e^{-z} \ln \pi \rho d z - \sum_{m=1}^{k-1} \frac{(\pi D^2 \rho)^m}{m!}  \\ 
    & ~~~~ \cdot  e^{-\pi D^2 \rho} \ln D  \Bigg| ^\infty_0 + \sum_{m=1}^{k-1} \int_{0}^{\infty} \frac{(\pi \rho D^2)^m}{m!} e^{-\pi \rho D^2}  \mbox{d} \left(\ln D\right)\\
    &  =  -\frac{\gamma}{2}-\frac{\ln \pi \rho}{2} + \sum_{m=1}^{k-1} \int_{0}^{\infty} \frac{(\pi \rho)^m}{m!} D^{2m-1} e^{-\pi \rho D^2}  \mbox{d} D \\
    & = -\frac{\gamma}{2}-\frac{\ln \pi \rho}{2} + \sum_{m=1}^{k-1} \frac{1}{2m},
    \end{split}
    \end{equation}
    {{where $\gamma$ is the Euler-Mascheroni constant with numerical value of 0.577215664902... \mbox{\cite{Euler_constant}}.}} Substituting (\ref{eq_appendix_lemma_2_1}) into (\ref{eq_appendix_A_2}) and (\ref{eq_appendix_A_1}), Lemma~1 can be proved.
    \hfill \rule{4pt}{8pt}\\
    
    {{The lower bound abstracts wireless transmission rate with respect to physical layer parameters and the group load distributions. 
    		As the group load distribution results from the cooperative caching schemes, Lemma 1 can be applied to analyze the performance of cooperative caching while taking into account wireless transmission features.
    		The lower-bound approximation is quite accurate in high SINR region, and hence supports conservative analysis.
    		In practical systems, most users are guaranteed with high SINR due to the reliable communication requirement, which can be achieved through advanced interference mitigation techniques.
    		The accuracy of the lower-bound will be validated in the simulation section.}}
    Lemma~1 indicates that the file transmission rate of Group-$k$ varies with the allocated bandwidth $\phi_k$ and traffic load $\Omega_k$.  	
    For notation simplicity, the average file transmission rate is rewritten as
    \begin{equation}
    \mathds{E}[R_k] \approx  \frac{\phi_k}{\Omega_k} W \tau_k,
    \end{equation}
    where 
    \begin{equation}
    \label{eq_R_k_tilde}
    \tau_k \triangleq  \frac{ \rho}{\lambda} \left[ \log_2 \frac{P_\mathrm{T} (\pi \rho)^{\frac{\alpha}{2}}}{\sigma^2 + I_k} + \frac{\alpha}{2\ln 2} (\gamma - \sum\limits_{m=1}\limits^{k-1} \frac{1}{m}) \right],
    \end{equation}
    representing the average spectrum efficiency of the $k$th candidate SBS.		
    Specifically, the physical meaning of $W \tau_k$ is the average file transmission rate when all users are served by their $k$th candidate SBS with all available system bandwidth.
    $\tau_k$ is irrelevant with content placement and bandwidth allocation, whereas depends on the overall traffic load (i.e., $\lambda/\rho$) and network resources (such as cell density, and transmitted SINR).
    Furthermore, $\tau_k$ decreases with $k$, due to high transmission distance and path loss.
    %This relationship demonstrates the tradeoff between caching diversity and spectrum efficiency.	 

    \subsection{Bandwidth Allocation Subproblem}
    
    {{Substituting the results of Lemma~1 into Eq.~(\ref{eq_delay_define}), the average file transmission delay can be rewritten as}}
    \begin{equation}
    \label{eq_delay_aver_2}
    \bar{D} = \left[ \sum\limits_{k=1}\limits^{K+1} \frac{{\Omega_k}^2}{W \tau_k \phi_k } \right] \bar{S}L + \bar{D}_\mathrm{BH} \Omega_{K+1}.
    \end{equation}
    For the given content placement $\{c_f\}$, $[\Omega_1,\Omega_2,\cdots,\Omega_{K+1}]$ can be derived based on Eqs.~(\ref{eq_hit_ratio_f}), (\ref{eq_hit_ratio_miss}), and (\ref{eq_group_ratio}).
    As a result, the average file transmission delay only varies with $\{\phi_k\}$ in Eq.~(\ref{eq_delay_aver_2}).
    The bandwidth allocation subproblem can be formulated as follows:
    \begin{subequations}
    	\begin{align}
    	\underset{\{\phi_k\}}{\min} ~~& \left[ \sum\limits_{k=1}\limits^{K+1} \frac{{\Omega_k}^2}{W \tau_k \phi_k } \right] \bar{S}L + \bar{D}_\mathrm{BH} \Omega_{K+1}, \\
    	\mbox{(SP1)}~~~~\mbox{s.t.}~~& \sum_{k=1}^{K+1} \phi_k \leq 1,\\
    	& \phi_k \geq 0,~~k=1,2,\cdots,K+1.
    	\end{align}
    \end{subequations}
    (SP1) is a convex optimization with respect to $\{\phi_k\}$, since $\frac{\partial \bar{D}^2}{\partial^2 \phi_k} = \frac{\Omega_k^2}{2 W \tau_k {\phi_k}^3} > 0$. 
    The optimal bandwidth allocation is given as Proposition~1, obtained with the Lagrange multiplier method.\\
    
    \textbf{Proposition 1.} For the given content placement, the optimal bandwidth allocation is given as
    \begin{equation}
    \label{eq_opt_band}
    \hat{\phi}_k = \frac{\frac{\Omega_k}{\sqrt{\tau_k}}}{\sum_{j=1}^{K+1} \frac{\Omega_j}{\sqrt{\tau_j}}}.
    \end{equation}
    
    %\emph{Proof.} Please refer to Appendix~\ref{appendix_proposition_1}.	
    \emph{Proof.} The Lagrange function of bandwidth allocation subproblem is given by
    \begin{equation}
    \begin{split}
    & ~~G(\phi_1,\cdots,\phi_{K+1})\\
    = & \sum\limits_{k=1}\limits^{K+1} \frac{{\Omega_k}^2 \bar{S}L}{W \phi_k \tau_k} \! + \! \bar{D}_\mathrm{BH} \Omega_{K\!+\!1} \!+\! \xi_0 \left(\! \sum_{k=1}^{K+1} \phi_k \!-\! 1 \! \right) \!-\! \sum_{k=1}^{K+1} \xi_k \phi_k,					
    \end{split}
    \end{equation}
    where $\xi_0 \geq 0$ is the Lagrange multiplier for constraint $\sum_{k=1}^{K+1} \phi_k \leq 1 $, $\xi_k \geq 0$ is the multiplier for constraint $\phi_k \geq 0$ and $k=1,2,\cdots, K+1$.
    Taking the derivative of the Lagrange function, the optimal solution should satisfy
    \begin{equation}
    \frac{\partial G(\phi_1,\cdots,\phi_{K+1})}{\partial \phi_k} = -  \frac{{\Omega_k}^2 \bar{S} L}{ W \tau_k {\phi_k}^2} + \xi_0 - \sum_{k=1}^{K+1} \xi_k = 0,
    \end{equation}
    which is equivalent to 
    \begin{equation}
    \phi_k = \Omega_k\sqrt{\frac{\bar{S}L}{W \tau_k (\xi_0 - \sum_{k=1}^{K+1}\xi_k)}}.
    \end{equation}
    {{Due to the complementary slackness of a constraint and its optimal Lagrange multiplier, $\phi_{k}>0$ implies $\mu_{k}=0$, $\forall k$. }}
    As a result, the optimal bandwidth allocation should also satisfy $\xi_0 \left(\sum_{k=1}^{K+1}\phi_k-1\right) = 0$, i.e.,
    \begin{equation}
    {\xi_0} \left(\sum_{k=1}^{K+1} \Omega_k\sqrt{\frac{\bar{S}L}{W \tau_k \xi}} - 1 \right) = 0.
    \end{equation}
    Therefore, $\sqrt{\xi_0} = \sum_{k=1}^{K+1} \Omega_k\sqrt{\frac{\bar{S}L}{W \tau_k }}$, and Proposition~1 is thus proved.	
    \hfill \rule{4pt}{8pt}\\

    %\subsection{Tradeoff between Content Diversity and Spectrum Efficiency}
    
    Substitute (\ref{eq_opt_band}) into (\ref{eq_delay_aver_2}), the average file transmission delay can be rewritten as
    \begin{equation}
    \small
    \label{eq_delay_detailed}
    \begin{split}
    & \bar{D} = \left(\sum_{k=1}^{K+1} \frac{\Omega_k}{\sqrt{\tau_k}} \right)^2 \frac{\bar{S} L}{W} + \bar{D}_\mathrm{BH} \Omega_{K+1} \\
    & = \left(\sum_{k=1}^{K} \frac{\Omega_k}{\sqrt{\tau_k}} \!+\! \frac{1\!-\!\sum_{k=1}^{K}\Omega_k}{\sqrt{\tau_1}} \right)^2 \frac{\bar{S} L}{W} \!+\! \bar{D}_\mathrm{BH} \left(1\!-\!\sum_{k=1}^{K}\Omega_k\right)\\
    & = \left( \sum_{k=2}^{K} \left( \frac{1}{\sqrt{\tau_k}} \!-\! \frac{1}{\sqrt{\tau_1}} \right) \Omega_k \!+\! \frac{1}{\sqrt{\tau_1}}\right)^2 \frac{\bar{S} L}{W} \!+\! \bar{D}_\mathrm{BH} \left(1\!-\!\sum_{k=1}^{K}\Omega_k\right).
    \end{split}
    \end{equation}
    The two parts can be interpreted as the average wireless transmission delay and backhaul delay, respectively.
    Notice that placing more diverse files in cache increases the content hit ratio $\Omega_k$ ($k=1,2,\cdots,K$).
    Based on Eq.~(\ref{eq_delay_detailed}), the average backhaul delay decreases, whereas the average wireless transmission delay increases with $\Omega_k$, where $k=2,3,\cdots,K$.
    This result reflects the tradeoff between content diversity gain and spectrum efficiency degradation under cooperative caching, which is influenced by the cache size.
    Specifically, larger cluster size brings higher content diversity gain, but results in lower spectrum efficiency.
    Take derivative of Eq.~(\ref{eq_delay_detailed}) with respect to $\Omega_k$:
    \begin{equation}
    \label{eq_derivative_D}
    \begin{split}
    \frac{\partial \bar{D}}{\partial \Omega_k} = & \frac{2 \bar{S} L}{W} \left( \sum_{k=2}^{K} \left( \frac{1}{\sqrt{\tau_k}} - \frac{1}{\sqrt{\tau_1}} \right) \Omega_k + \frac{1}{\sqrt{\tau_1}} \right)\\
    & \cdot \left( \frac{1}{\sqrt{\tau_k}} - \frac{1}{\sqrt{\tau_1}} \right) - \bar{D}_\mathrm{BH},				
    \end{split}
    \end{equation}
    where $k=2,3,\cdots,K$.
    If $\frac{\partial \bar{D}}{\partial \Omega_k} \leq 0$, the backhaul delay dominants, and we should enhance content diversity to reduce average file transmission delay.
    Otherwise, the wireless transmission delay dominants, and using the $k$th candidate SBS may even increase average transmission delay due to the large path loss.	
    To balance the gain of content diversity and degradation of spectrum efficiency, $\frac{\partial \bar{D}}{\partial \Omega_k} \leq 0$ should be always guaranteed, and a sufficient condition is given in Proposition~2.\\
    
    \textbf{Proposition 2}.
    If 
    \begin{equation}
    \label{eq_proposition_2}
    \frac{2 \bar{S} L}{W \sqrt{\tau_K}}\left( \frac{1}{\sqrt{\tau_K}} - \frac{1}{\sqrt{\tau_1}} \right) \leq \bar{D}_\mathrm{BH},
    \end{equation}
    $\frac{\partial \bar{D}}{\partial \Omega_k} \leq 0$ can be guaranteed $\forall \Omega_k \in [0,1]$, where $k=2,3,\cdots,K$.
    
    %\emph{Proof.} Please refer to Appendix~\ref{appendix_proposition_2}.
    \emph{Proof.} As $\tau_{K+1} = \tau_1$ and $\tau_{K}\leq \tau_{k}$ (for $k=1,2,\cdots,K+1$), we have
    \begin{equation}
    \label{eq_appendix_pro_2_1}
    \begin{split}
    & \sum_{k=2}^{K} \left( \frac{1}{\sqrt{\tau_k}} - \frac{1}{\sqrt{\tau_1}} \right) \Omega_k + \frac{1}{\sqrt{\tau_1}}\\
    & = \sum_{k=1}^{K+1} \left( \frac{1}{\sqrt{\tau_k}} - \frac{1}{\sqrt{\tau_1}} \right) \Omega_k + \frac{1}{\sqrt{\tau_1}} \sum_{k=1}^{K+1} \Omega_k  \\
    & = \sum_{k=1}^{K+1} \frac{1}{\sqrt{\tau_k}} \Omega_k \leq  \frac{1}{\sqrt{\tau_{K}}}.
    \end{split}
    \end{equation}
    Furthermore, 
    \begin{equation}
    \label{eq_appendix_pro_2_2}
    \frac{1}{\sqrt{\tau_k}} - \frac{1}{\sqrt{\tau_1}} \leq \frac{1}{\sqrt{\tau_K}} - \frac{1}{\sqrt{\tau_1}},
    \end{equation}
    Substitute Eqs.~(\ref{eq_appendix_pro_2_1}) and (\ref{eq_appendix_pro_2_2}) into Eq.~(\ref{eq_derivative_D}):
    \begin{equation}
    \frac{\partial \bar{D}}{\partial \Omega_k} \leq \frac{2 \bar{S} L}{W} \frac{1}{\sqrt{\tau_{K}}} \left( \frac{1}{\sqrt{\tau_K}} - \frac{1}{\sqrt{\tau_1}} \right) - \bar{D}_\mathrm{BH}.
    \end{equation}  
    Therefore, Eq.~(\ref{eq_proposition_2}) provides a sufficient condition to $\frac{\partial \bar{D}}{\partial \Omega_k} \leq 0$, and Proposition~2 is thus proved.
    \hfill \rule{4pt}{8pt}\\
    
    Notice that the left part of Eqn.~(\ref{eq_proposition_2}) can be interpreted as the maximal degradation of spectrum efficiency when the cluster size is $K$\footnote{``Eqn''. is short for ``inequation'', and ``Eq.'' is short for ``Equation''}..
    Therefore, Proposition~2 indicates that the degradation of spectrum efficiency should be no larger than the backhaul delay.
    Otherwise, the delay of remote file fetching is even smaller than the delay of fetching files from SBS $B_K$, and $B_K$ should not be involved into cluster.
    Proposition~2 offers a guideline for SBS clustering.
    As the left part of Eqn.~(\ref{eq_proposition_2}) (i.e., the spectrum efficiency degradation) increases with $K$, Eqn.~(\ref{eq_proposition_2}) constraints the maximal cluster size.
    Furthermore, the maximal cluster size depends on network parameters.
    For example, larger backhaul delay indicates larger cluster size.
    In addition, the cluster size can increase in dense networks, as $\tau_K$ increases with shorter transmission distance.
%%%%%%%%%%%%%%%%%%%%%%%%%%%%%%%%%%%%%%%%%%%%%%%%%%%%%%%%%%%%%%%%%%%%%%%%%%%%%%%%%%%%%%%%%%%%%%%%%%%
\section{Content Placement with Optimal Bandwidth Allocation}
	\label{sec_submodular}
	{{Based on the result of optimal bandwidth allocation, the content placement subproblem can be formulated as follows}}:
	\begin{subequations}
		\begin{align}
		\underset{\{c_f\}}{\min} ~~& \frac{\bar{S} L}{W} \left(\sum_{k=1}^{K+1} \frac{\Omega_k}{\sqrt{\tau_k}} \right)^2   + \bar{D}_\mathrm{BH} \Omega_{K+1}, \\
		\mbox{(SP2)}~~~~\mbox{s.t.}~~& \sum_{f=1}^{F} c_f  \leq C,\\
		& c_f\in\{0,1,\cdots,s_f\}, ~~\forall f\in\mathcal{F}.
		\end{align}
	\end{subequations}
	The complexity of this integer programing problem increases exponentially with the number of files and file size.
	In addition, the average delay is piecewise with respect to caching placement based on Eqs.~(\ref{eq_hit_ratio_f}), (\ref{eq_hit_ratio_miss}) and (\ref{eq_group_ratio}), posing additional challenges.
	Therefore, a sub-optimal algorithm with low complexity should be devised for practical network operations.
	%A widely adopted method is to approximate the integer decision variables to be continuous, and then apply convex optimization theories such as Lagrangian multiplier method.
	To this end, we first analyze the delay improvement by adding a segment into cache, and then design a greedy algorithm based on the submodular properties.
	
	\subsection{Delay Improvement with Segment Placement}
	
	When a segment is added into cache, the content hit ratio increases, influencing average file transmission delay.
	Suppose the number of segments cached in the SBS is $\mathcal{C}=\{c_1,c_2,\cdots,c_f,\cdots,c_F\}$, whose corresponding content hit ratio and group load distribution is $\{P_{k,f}\}$ and $\{\Omega_k\}$, respectively.
	Assume file-$h$ is not completely stored (i.e., $c_h < s_h$), and we add a segment of file-$h$ into cache.
	Accordingly, the number of cached segments becomes $\mathcal{C}'=\{c'_1,c'_2,\cdots,c'_f,\cdots,c'_F\}$, where $c'_h=c_h+1$, $c'_f=c_f$, $\forall f\in \mathcal{F}$ and $f\neq h$. 
	Denote by  $\{P'_{k,f}\}$ and $\{\Omega'_k\}$ the updated content hit ratio and group load distribution, respectively.
	To evaluate the delay performance improved by caching the segment, we define $V(\mathcal{C},h)$ as the marginal gain:
	\begin{equation}
	\label{eq_marginal_gain_def}
	V(\mathcal{C},h) = \bar{D}\Big|_{\mathcal{C}} - \bar{D}\Big|_{\mathcal{C}'},
	\end{equation} 
	where $\bar{D}\Big|_{\mathcal{C}}$ and $\bar{D}\Big|_{\mathcal{C}'}$ denote the average file transmission delays when the cached segment set is $\mathcal{C}$ and $\mathcal{C}'$, respectively.	
	Substitute Eq.~(\ref{eq_delay_detailed}) into Eq.~(\ref{eq_marginal_gain_def}), and the marginal gain is given by
	\begin{equation}
	\label{eq_marginal}
	\begin{split}
	V(\mathcal{C},h) =  & \bar{D}_\mathrm{BH} (\Omega_{K+1}-\Omega'_{K+1})\\
	& + \frac{\bar{S} L}{W} \left(\sum_{k=1}^{K+1} \frac{\Omega_k+\Omega'_k}{\sqrt{\tau_k}} \right) \left(\sum_{k=1}^{K+1} \frac{\Omega_k - \Omega'_k}{\sqrt{\tau_k}} \right) ,				
	\end{split}
	\end{equation}
	indicating that the key factor of the marginal gain is the variation of group load distribution.
	
	Denote by $\delta_k = \Omega'_k - \Omega_k$ and $\Delta = [\delta_1, \delta_2, \cdots, \delta_{K+1}]$, i.e., the variations of group load distribution.  
	The group load distribution depends on the content hit ratio, according to Eq.~(\ref{eq_group_ratio}).
	For the other files except $h$, the number of cached segment remains the same, and thus the content hit ratio does not change, i.e., $P'_{k,f}=P_{k,f}$ for $f\neq h$.
	Therefore, the variations of group load distribution is given by $\delta_k = q_h (P'_{k,h}-P_{k,h})$ for $k=1,2,\cdots,K+1$.
	As the content hit ratio is piecewise with respect to the SBS index $k$ according to Eq.~(\ref{eq_hit_ratio_f}) and (\ref{eq_hit_ratio_miss}), $\delta_k$ also has a piecewise structure. 
	The result of $\delta_k$ can be classified into the following two cases, depending on the number of cached segments  $c_h$ \footnote{Assume ${s_f/K}$ is an integer without losing generality.}.	\\
	
	\textbf{Case 1.} If $c_h \leq \frac{s_h}{K} - 1$, the variation of group load distribution is given by
	\begin{equation}
	\label{eq_case_1}
	\delta_k = \left\{ \begin{array}{ll}
	\frac{q_h}{s_h},~~&~~k=1,2,\cdots,K,\\
	-\frac{K q_h}{s_h} ,~~&~~k=K+1.
	\end{array} \right.
	\end{equation}
	
	\textbf{Case 2.} If $c_h \geq \frac{s_h}{K}$, the variation of group load distribution is given by
	\begin{equation}
	\label{eq_case_2}
	\left\{ \begin{array}{ll}
	\delta_k = \frac{q_h}{s_h},~~&~~k=1,2,\cdots,\check{K},\\
	\delta_k < 0,~~&~~k=\check{K}+1,\cdots,\hat{K},\\
	\delta_k = 0,~~&~~k=\hat{K}+1,\cdots,K+1,
	\end{array} \right.
	\end{equation}
	where
	%		\begin{equation}
	%			\sum_{k=1}^{\hat{K}-1}\delta_k = 0,
	%		\end{equation}
	$\sum_{k=1}^{\hat{K}}\delta_k = 0$, $\check{K} = \lfloor \frac{s_h}{c_h+1} \rfloor$, and $\hat{K}=\lceil \frac{s_h}{c_h} \rceil$.\\
	
	When the number of cached segments is small, extra segments need to be fetched from remote servers for file decoding.
	Accordingly, adding a new segment into cache can reduce content miss ratio, which is the situation of Case 1.
	However, when the number of cached segments exceeds some threshold, the requested file can be decoded by downloading files locally from the candidate SBSs, i.e., no content miss in Case 2.
	Therefore, adding a new segment into cache no longer improves total content hit ratio, but users can finish file downloading from closer SBSs with higher transmission rate. 
	For example, suppose there are 6 candidate SBSs and consider a file-$h$ requiring 100 segments to decode.
	If each SBS caches 24 segments, users need to download segments from SBS $B_1$-$B_5$ for file decoding. 
	If each SBS caches 25 segments, users can finish file decoding only from SBS $B_1$-$B_4$.
	In this case, $\check{K}=4$, $\hat{K}=5$, and $\Delta = [\frac{1}{200}, \frac{1}{200}, \frac{1}{200}, \frac{1}{200}, -\frac{1}{50}, 0, 0]$ if the popularity is $q_h=0.5$.
	
	The tradeoff between content diversity and spectrum efficiency influences the marginal gain differently in the two cases.
	In Case 1, adding a new segment enhances content diversity as well as content hit ratio.
	Accordingly, the backhaul transmission delay can be reduced, but the wireless transmission delay increases due to degraded spectrum efficiency.
	In Case 2, adding a new segment cannot enhance content diversity, and brings no backhaul delay improvement.
	However, wireless transmission delay can be reduced with improved spectrum efficiency, since users can download file from BSs in proximity.	
	
	\subsection{Submodular Property Analysis}
	
	In what follows, we analyze the marginal gain of content placement with respect to $\mathcal{C}$ and $h$. 
	Specifically, the marginal gain can be proved to have monotone submodular properties under the condition of Proposition~2, given as Proposition~3.\\
	
	\textbf{Proposition 3.} For any two feasible cache placement $\mathcal{C}=\{c_1,\cdots,c_F\}$ and $\mathcal{C}'=\{c'_1,\cdots,c'_F\}$ satisfying $c_f \leq c'_f$ ($\forall f\in \mathcal{F}$), we have $V(\mathcal{C}, h) \geq V(\mathcal{C}',h) > 0$ under the condition of Proposition 2, where $c_h \leq c'_h < s_h$.
	
	%\emph{Proof.} Please refer to Appendix~\ref{appendix_proposition_3}.
	\emph{Proof.} The proof can be conducted in two steps: 1) the marginal gain is positive, i.e., $V(\mathcal{C},h)>0$; and 2) the marginal gain decreases with the number of cached contents, i.e., $V(\mathcal{C}',h)<V(\mathcal{C},h)$.
	For the proof of step 2), we prove $V(\mathcal{C}',h)<V(\mathcal{C},h)$ under a special case: suppose $\mathcal{C}$ is the initial cache placement where $c_h\leq s_h-2$, $\mathcal{C}'$ is the cache placement when a segment of file-$h$ is added, and $\mathcal{C}''$ is the cache placement when two segments of file-$h$ are added.
	For the general case of $c'_f \geq c_f$, $V(\mathcal{C}',h)<V(\mathcal{C},h)$ can be proved through recursion.
	Furthermore, the proof needs to be conducted under both cases, respectively.
	\vspace{1.5mm}\\
	\textbf{(1) $V(\mathcal{C},h)>0$ in Case 1:}
	
	Substituting Eq.~(\ref{eq_case_1}) into (\ref{eq_marginal}), the marginal value can be rewritten as 
	\begin{equation}
	\label{eq_marginal_case_1}
	\begin{split}
	& ~~ V(\mathcal{C},h) \\
	= &~K \bar{D}_\mathrm{BH} \frac{q_h}{s_h} - \frac{\bar{S} L}{W} \left(\sum_{k=1}^{K+1}\frac{\Omega_k}{\sqrt{\tau_k}} + \sum_{k=1}^{K+1}\frac{\Omega'_k}{\sqrt{\tau_k}} \right) \left( \sum_{k=1}^{K+1}\frac{\delta_k}{\sqrt{\tau_k}} \right)\\
	= &~K \bar{D}_\mathrm{BH} \frac{q_h}{s_h} - \frac{\bar{S} L q_h}{W s_h}  \left(\sum_{k=1}^{K\!+\!1}\frac{\Omega_k\!+\!\Omega'_k}{\sqrt{\tau_k}} \right)  \sum_{k=1}^{K}\left(\frac{1}{\sqrt{\tau_k}} \! -\! \frac{1}{\sqrt{\tau_{K\!+\!1}}}\right)
	\end{split}				
	\end{equation}
	As $\tau_{K+1}=\tau_1 > \tau_2 > \cdots>\tau_K$, we have
	\begin{equation}
	\label{eqn_rate_1}
	\sum_{k=1}^{K}\left( \frac{1}{\sqrt{\tau_k}} - \frac{1}{\sqrt{\tau_1}} \right) < K \left( \frac{1}{\sqrt{\tau_K}} - \frac{1}{\sqrt{\tau_1}} \right),
	\end{equation} 	
	and 
	\begin{equation}
	\label{eqn_rate_2}
	\frac{2}{\sqrt{\tau_K}} > \sum_{k=1}^{K+1}\frac{\Omega_k}{\sqrt{\tau_k}} + \sum_{k=1}^{K+1}\frac{\Omega'_k}{\sqrt{\tau_k}},
	\end{equation}
	due to $\sum_{k=1}^{K+1} \Omega_k = \sum_{k=1}^{K+1} \Omega'_k=1$.
	According to (\ref{eq_proposition_2}),
	\begin{equation}
	\begin{split}
	& K \bar{D}_\mathrm{BH} \geq 2 K \frac{\bar{S} L}{W} \frac{1}{\sqrt{\tau_K}} \left( \frac{1}{\sqrt{\tau_K}} - \frac{1}{\sqrt{\tau_1}} \right)  \\
	& > \frac{\bar{S} L}{W} \left( \sum_{k=1}^{K+1}\frac{\Omega_k}{\sqrt{\tau_k}} + \sum_{k=1}^{K+1}\frac{\Omega'_k}{\sqrt{\tau_k}} \right) \sum_{k=1}^{K} \left( \frac{1}{\sqrt{\tau_k}} - \frac{1}{\sqrt{\tau_1}} \right).
	\end{split}
	\end{equation} 
	Hence, $V(\mathcal{C},h)>0$.
	\vspace{1.5mm}\\
	\textbf{(2) $V(\mathcal{C},h)>0$ in Case 2:}
	
	Substituting Eq.~(\ref{eq_case_2}) into (\ref{eq_marginal}), the marginal value can be rewritten as 
	\begin{equation}
	\label{eq_marginal_case_2}
	V(\mathcal{C},h) = - \frac{\bar{S} L}{W} \left(\sum_{k=1}^{K+1}\frac{\Omega_k}{\sqrt{\tau_k}} + \sum_{k=1}^{K+1}\frac{\Omega'_k}{\sqrt{\tau_k}} \right) \left( \sum_{k=1}^{\hat{K}-1}\frac{\delta_k}{\sqrt{\tau_k}} \right).
	\end{equation}
	As $\delta_1=\delta_2=\cdots=\delta_{\check{K}}={q_h/s_h} > 0 > \delta_{\check{K}+1} \geq \cdots \geq \delta_{\hat{K}}$ and $\tau_1>\tau_2>\cdots>\tau_{\hat{K}}$, we have
	\begin{equation}
	\label{eq_eqn_negative_1}
	\begin{split}
	& \sum_{k=1}^{\hat{K}}\frac{\delta_k}{\sqrt{\tau_k}} = \sum_{k=1}^{\check{K}}\frac{\delta_k}{\sqrt{\tau_k}} + \sum_{k=\check{K}+1}^{\hat{K}}\frac{\delta_k}{\sqrt{\tau_k}} \\
	< & \sum_{k=1}^{\check{K}}\frac{\delta_k}{\sqrt{\tau_{\check{K}}}} + \sum_{k=\check{K}+1}^{\hat{K}}\frac{\delta_k}{\sqrt{\tau_{\check{K}}}}=0.				
	\end{split}
	\end{equation}
	Hence, $V(\mathcal{C},h)>0$.
	\vspace{1.5mm}\\
	\textbf{(3) $V(\mathcal{C}',h)<V(\mathcal{C},h)$ in Case 1:}
	
	Denote by $\{\Omega_k\}$, $\{\Omega'_k\}$ and $\{\Omega''_k\}$ the group load distributions, corresponding to cache placement is $\mathcal{C}$, $\mathcal{C}'$ and $\mathcal{C}''$, respectively.
	Denote by $\delta_k = \Omega'_k-\Omega_k$, and $\delta'_k = \Omega''_k-\Omega'_k$, for $k=1,\cdots,K+1$.
	According to Eq.~(\ref{eq_marginal_case_1}) and $\tau_{K+1}=\tau_1$, we have
	\begin{equation}
	\begin{split}
	& V(\mathcal{C}',h) - V(\mathcal{C},h)  \\
	=  & \frac{\bar{S} L}{W} \frac{q_h}{s_h} \left(\sum_{k=1}^{K+1}\frac{\Omega_k-\Omega''_k}{\sqrt{\tau_k}}\right)  \sum_{k=1}^{K}\left(\frac{1}{\sqrt{\tau_k}} - \frac{1}{\sqrt{\tau_{1}}}\right).
	\end{split}
	\end{equation}
	As $\Omega_k-\Omega''_k=-{2q_h/s_h}$ for $k=1,\cdots,K$ and $ \Omega_{K+1}-\Omega''_{K+1} =  {2K q_h/s_h}$, we have
	\begin{equation}
	\sum_{k=1}^{K+1}\frac{\Omega_k-\Omega''_k}{\sqrt{\tau_k}} = -2\frac{q_h}{s_h}  \sum_{k=1}^{K+1}\left(\frac{1}{\sqrt{\tau_k}}-\frac{1}{\sqrt{\tau_{1}}}\right).
	\end{equation}
	Therefore, 
	\begin{equation}
	\begin{split}
	& V(\mathcal{C}',h) - V(\mathcal{C},h) \\
	= & - \frac{2 \bar{S} L}{W} \left(\frac{q_h}{s_h}\right)^2 \left[\sum_{k=1}^{K+1}\left(\frac{1}{\sqrt{\tau_k}}-\frac{1}{\sqrt{\tau_{1}}}\right)\right]^2 <0.
	\end{split}				
	\end{equation}
	\vspace{1.5mm}\\
	\textbf{(4) $V(\mathcal{C}',h)<V(\mathcal{C},h)$ in Case 2:}
	
	According to Eq.~(\ref{eq_marginal_case_2}), we have
	\begin{equation}
	\begin{split}
	&~~V(\mathcal{C}', h) - V(\mathcal{C}, h) \\ 
	= & \frac{\bar{S} L}{W} \left[ \left(\sum_{k=1}^{K+1}\frac{\Omega_k}{\sqrt{\tau_k}} + \sum_{k=1}^{K+1}\frac{\Omega'_k}{\sqrt{\tau_k}} \right)  \left( \sum_{k=1}^{\hat{K}}\frac{\delta_k}{\sqrt{\tau_k}} \right) \right. \\
	= & \left. - \left(\sum_{k=1}^{K+1}\frac{\Omega'_k}{\sqrt{\tau_k}} +  \sum_{k=1}^{K+1}\frac{\Omega''_k}{\sqrt{\tau_k}} \right) \left( \sum_{k=1}^{\hat{K}'}\frac{\delta'_k}{\sqrt{\tau_k}} \right)\right],					
	\end{split}
	\end{equation}
	where $\hat{K}=\lceil\frac{s_h}{c_h}\rceil$ and $\hat{K}'= \lceil\frac{s_h}{c_h+1}\rceil$.
	Notice that $\Omega''_k-\Omega_k$ corresponds to the variations of group load distributions when two segments of file-$h$ are added, which has a similar structure of $\delta_k$:
	\begin{equation}
	\left\{ \begin{array}{ll}
	\Omega''_k-\Omega_k = \frac{2 q_h}{s_h},~~&~~k=1,2,\cdots,\check{K}'',\\
	\Omega''_k-\Omega_k < 0,~~&~~k=\check{K}''+1,\cdots,\hat{K}'',\\
	\Omega''_k-\Omega_k = 0,~~&~~k=\hat{K}''+1,\cdots,K+1,
	\end{array} \right.
	\end{equation}
	%		and
	%			\begin{equation}
	%			\sum_{k=1}^{\hat{K}-1}(\Omega''_k-\Omega_k) = 0,
	%			\end{equation}
	where $\sum_{k=1}^{\hat{K}''}(\Omega''_k-\Omega_k) = 0$, $\check{K}'' = \lfloor \frac{s_h}{c_h+2} \rfloor$, and $\hat{K}''=\lceil \frac{s_h}{c_h} \rceil$.
	Due to the same reason of Eqn.~(\ref{eq_eqn_negative_1}), 
	\begin{equation}
	\sum_{k=1}^{K+1}\frac{\Omega''_k-\Omega_k}{\sqrt{\tau_k}} <0.
	\end{equation}
	Therefore,
	\begin{equation}
	\label{eqn_appendix_E_3}
	\sum_{k=1}^{K+1}\frac{\Omega_k}{\sqrt{\tau_k}} + \sum_{k=1}^{K+1}\frac{\Omega'_k}{\sqrt{\tau_k}} > \sum_{k=1}^{K+1}\frac{\Omega'_k}{\sqrt{\tau_k}}+ \sum_{k=1}^{K+1}\frac{\Omega''_k}{\sqrt{\tau_k}} > 0.
	\end{equation}
	
	In what follows, we prove $\sum_{k=1}^{\hat{K}}\frac{\delta_k}{\sqrt{\tau_k}} < \sum_{k=1}^{\hat{K}'}\frac{\delta'_k}{\sqrt{\tau_k}} < 0$.		
	As $c_h \geq \frac{s_h}{K}$ in Case 2, we have $\frac{s_h}{c_h} - \frac{s_h}{c_h+1} = \frac{s_h}{(c_h+1)c_h} < \frac{K^2}{s_h}$.
	In practical systems, the number of candidate SBSs $K$ (usually smaller than 10) is generally much smaller than $s_h$ (usually hundreds or thousands).
	Thus, we assume $\frac{s_h}{c_h} - \frac{s_h}{c_h+1} \leq 1$.
	Therefore, $\hat{K}' \leq \hat{K}\leq \hat{K}'+1$, and $\check{K} \leq \hat{K} \leq \check{K} + 2$.
	Similarly, $\check{K}' \leq \check{K}\leq \check{K}'+1$, $\check{K}' \leq \hat{K}' \leq \check{K}' + 2$, and $\check{K}' \leq \hat{K} \leq \check{K}' + 2$.
	
	If $\check{K}=\check{K}'$, we have $\hat{K}'=\check{K}'+1$, $\delta_k=\delta'_k={q_h/s_h}$ for $k=1,2,\cdots,\check{K}$, $\delta_{\check{K}+1}<0$, $\delta_{\check{K}+2} \leq 0$, $\delta'_{\check{K}+1}<0$, $\delta'_{\check{K}+2}=0$, and $\delta_k=\delta'_k=0$ for $k=\check{K}+3,\cdots,K$. Thus,
	\begin{equation}
	\label{eq_eqn_negative_2}
	\begin{split}
	& \sum_{k=1}^{\hat{K}}\frac{\delta_k}{\sqrt{\tau_k}} = \sum_{k=1}^{\check{K}}\frac{\delta_k}{\sqrt{\tau_k}} + \frac{\delta_{\check{K}+1}}{\sqrt{\tau_{\check{K}+1}}} + \frac{\delta_{\check{K}+2}}{\sqrt{\tau_{\check{K}+2}}},			
	\end{split}
	\end{equation}
	and 
	\begin{equation}
	\label{eq_eqn_negative_3}
	\begin{split}
	& \sum_{k=1}^{\hat{K}'}\frac{\delta'_k}{\sqrt{\tau_k}} = \sum_{k=1}^{\check{K}}\frac{\delta'_k}{\sqrt{\tau_k}} + \frac{\delta'_{\check{K}+1}}{\sqrt{\tau_{\check{K}+1}}}.			
	\end{split}
	\end{equation}
	As $\sum_{k=1}^{\check{K}+2}\delta_k = \sum_{k=1}^{\check{K}+1}\delta'_k =0$, $\delta_{\check{K}+1}+\delta_{\check{K}+1}=\delta'_{\check{K}+1}=-{ \check{K} q_h/s_h} < 0$, we further have
	\begin{equation}
	\label{eqn_appendix_E_2}
	\begin{split}					
	& \sum_{k=1}^{\hat{K}}\frac{\delta_k}{\sqrt{\tau_k}} - \sum_{k=1}^{\hat{K}'}\frac{\delta'_k}{\sqrt{\tau_k}} \\
	= & \frac{\delta_{\check{K}+1}}{\sqrt{\tau_{\check{K}+1}}} + \frac{\delta_{\check{K}+2}}{\sqrt{\tau_{\check{K}+2}}} - \frac{\delta_{\check{K}+1}+\delta_{\check{K}+2}}{\sqrt{\tau_{\check{K}+1}}} \\
	= & \delta_{\check{K}+2}\left(\frac{1}{\sqrt{\tau_{\check{K}+2}}} - \frac{1}{\sqrt{\tau_{\check{K}+1}}}\right) \leq 0,
	\end{split}
	\end{equation}
	since $\tau_{\check{K}+1} > \tau_{\check{K}+2}$.
	
	If $\check{K}>\check{K}'$, we have $\check{K}=\check{K}'+1$, $\hat{K}' \leq \check{K}'+2$, and $\hat{K} = \check{K}+1$.
	Thus, $\delta_k=\delta'_k={q_h/s_h}$ for $k=1,2,\cdots,\check{K}'$, $\delta_{\check{K}'+1}={q_h/s_h}$,  $\delta_{\check{K}'+2}<0$,  $\delta'_{\check{K}'+1}<0$, $\delta'_{\check{K}'+2}\leq 0$, and $\delta_k=\delta'_k=0$ for $k=\check{K}'+3,\cdots,K$.
	Accordingly,
	\begin{equation}
	\label{eqn_appendix_E_1}
	\begin{split}					
	& \sum_{k=1}^{\hat{K}}\frac{\delta_k}{\sqrt{\tau_k}} - \sum_{k=1}^{\hat{K}'}\frac{\delta'_k}{\sqrt{\tau_k}} \\
	= & \frac{\delta_{\check{K}'+1}}{\sqrt{\tau_{\check{K}'+1}}} + \frac{\delta_{\check{K}'+2}}{\sqrt{\tau_{\check{K}'+2}}} - \frac{\delta'_{\check{K}'+1}}{\sqrt{\tau_{\check{K}'+1}}}- \frac{\delta'_{\check{K}'+2}}{\sqrt{\tau_{\check{K}'+2}}}  \\
	< & \frac{\delta_{\check{K}'+1}+\delta_{\check{K}'+2}}{\sqrt{\tau_{\check{K}'+2}}} - \frac{\delta'_{\check{K}'+1}+\delta'_{\check{K}'+2}}{\sqrt{\tau_{\check{K}'+2}}} = 0,
	\end{split}
	\end{equation}
	since $\delta_{\check{K}'+1}+\delta_{\check{K}'+2} = \delta'_{\check{K}'+1}+\delta'_{\check{K}'+2} = - {\check{K}' q_h/s_h}$. 		
	With Eqns.~(\ref{eqn_appendix_E_2}) and (\ref{eqn_appendix_E_1}), we have $\sum_{k=1}^{\hat{K}}\frac{\delta_k}{\sqrt{\tau_k}} < \sum_{k=1}^{\hat{K}'}\frac{\delta'_k}{\sqrt{\tau_k}}$. Furthermore, we can prove $\sum_{k=1}^{\hat{K}}\frac{\delta_k}{\sqrt{\tau_k}} \!<\! 0$ and $ \sum_{k=1}^{\hat{K}'}\frac{\delta'_k}{\sqrt{\tau_k}} < 0 $ in the same way as Eqn.~(\ref{eq_eqn_negative_1}). Combining $\sum_{k=1}^{\hat{K}}\frac{\delta_k}{\sqrt{\tau_k}} < \sum_{k=1}^{\hat{K}'}\frac{\delta'_k}{\sqrt{\tau_k}} <0 $ with Eqn.~(\ref{eqn_appendix_E_3}), $V(\mathcal{C}',h) < V(\mathcal{C},h)$ can be proved in Case 2. 
	
	Therefore, $0< V(\mathcal{C}',h)< V(\mathcal{C},h)$ holds for both cases, and Proposition~3 can be proved.
	\hfill \rule{4pt}{8pt}\\
	
	\emph{Remark.} Proposition 3 reveals two facts: 1) adding a segment in cache can always improve the delay performance; and 2) the marginal gain of adding segments in cache decreases when more segments are stored in cache. 
	With this monotone submodular property, greedy algorithms can be designed with near-optimal performance guarantee.
	
	\begin{algorithm}[t!]
		\caption{The Proposed Greedy Algorithm} %\label{euclid}
		\begin{algorithmic}[1]
			\Require
			$\mathcal{F}$: file library;
			$\{q_f\}$: file popularity distribution; 
			$\{s_f\}$: file length;
			$C$: cache size of SBSs;
			$\rho$: SBS density;
			$K$: cluster size;
			$\bar{D}_\mathrm{BH}$: average backhaul delay; 
			\Ensure
			$\{c_f\}$: number of cached segments;
			$\{\hat{\phi}_f\}$: the optimal bandwidth allocation;
			\State Calculate $\tau_k$ based on Eq.~(\ref{eq_R_k_tilde}), for $k=1,2,\cdots,K+1$;
			\State Set $\mathcal{Y}_f=\emptyset$,~~$c_f=0$, $\forall f\in\mathcal{F}$;
			\State \textbf{while}~~$\sum_{f=1}^{F} c_f < C$~~\textbf{do}
			\State ~~~Calculate the group load distribution $\{\Omega_k\}$ with Eq.~(\ref{eq_load_distribution_new});
			\State ~~~$h^*=\arg \underset{h}{\max}~V(\{c_f\}, h)$, based on Eq.~(\ref{eq_marginal});
			\State ~~~$\mathcal{Y}_{h^*} \gets \mathcal{Y}_{h^*} \cup \{x_{c_{h^*}+1}\}$,~~$c_{h^*} = c_{h^*}+1$;
			\State \textbf{end while}
			\State Calculate $\{\hat{\phi}_f\}$ with Proposition~1;
			\State Return $\{c_f\}$ and $\{\hat{\phi}_f\}$;
			%			\EndProcedure
		\end{algorithmic}
	\end{algorithm}
	
	\subsection{Greedy Content Placement Algorithm}
	
	The content placement problem can be transferred into a submodular optimization problem.
	Denote by $\mathcal{X}_f$ a set of coded segments which can support the decoding of file-$f$, and $|\mathcal{X}_f|=s_f$.
	Denote by $\mathcal{Y}_f \subseteq \mathcal{X}_f$ the set of segments placed in cache, and $|\mathcal{Y}_f|=c_f$ in problem (SP2) \footnote{$|\cdot|$ denotes the cardinality of a set.}.	
	The content placement can be reformulated as (SP3).		
	\begin{subequations}
		\begin{align}
		\underset{\{\mathcal{Y}_f\}}{\max} ~~& \left[\! \frac{\bar{S} L}{W \tau_1}+\bar{D}_\mathrm{BH} \!\right] \!-\! \left[ \! \frac{\bar{S} L}{W} \left(\sum_{k=1}^{K+1} \frac{\Omega_k}{\sqrt{\tau_k}} \right)^2 \!+\! \bar{D}_\mathrm{BH} \Omega_{K+1} \! \right], \nonumber\\
		\mbox{(SP3)} &  ~~\mbox{s.t.}~~\sum_{f=1}^{F} |\mathcal{Y}_f|  \leq C,\\
		& ~~~~~~~\mathcal{Y}_f \subseteq \mathcal{X}_f, ~~\forall f\in\mathcal{F},
		\end{align}
	\end{subequations}
	where the group load distribution is given by
	\begin{equation}
	\label{eq_load_distribution_new}
	\Omega_k = \sum_{f=1}^{F} \frac{q_f}{s_f} \left[\min\left( k |\mathcal{Y}_f|, s_f \right) \!-\! \min\left( (k-1) |\mathcal{Y}_f|, s_f \right) \right],
	\end{equation}
	for $k=1,2,\cdots,K$, and $\Omega_{K+1} = 1-\sum_{k=1}^{K} \Omega_k$.
	The objective function represents the reduced delay with cooperative caching, where $ \frac{\bar{S} L}{W \tau_1}+\bar{D}_\mathrm{BH}$ denotes the average delay without content caching.
	A greedy content placement algorithm is proposed to solve problem (SP3) as described in Algorithm~1, where $\mathcal{Y}_f$ denotes the set of segments in cache, $c_f$ is the number of segments cached for file-$f$.
	Each time a segment is placed in cache until the caching capacity $C$ is achieved, and the one providing maximal marginal gain is always selected to minimize average delay. 
	The reformulated content placement problem (SP3) can be proved to be monotone submodular based on the result of Proposition~3.
	Therefore, the proposed greedy algorithm can achieve $(1-\frac{1}{e})$-optimality \cite{Krause12_submodular}.
	The computation complexity of the proposed greedy algorithm is $\mathcal{O}(F C)$.
	Specifically, $\mathcal{O}(F)$ denotes the computation complexity to find the file segment which can bring the maximal marginal gain, and $\mathcal{O}(C)$ corresponds to the process that $C$ segments are selected for caching.
%%%%%%%%%%%%%%%%%%%%%%%%%%%%%%%%%%%%%%%%%%%%%%%%%%%%%%%%%%%%%%%%%%%%%%%%%%%%%%%%%%%%%%%%%%%%%%%%%%%
\section{Simulation Results}
    \label{sec_simulation}
    \begin{table}[!t]
    	\caption{Simulation parameters \cite{Liu16_EE_cache_JSAC}}
    	\label{tab_parameter}
    	\centering
    	\begin{tabular}{cccc}
    		\hline
    		\hline
    		Parameter & Value & Parameter & Value \\
    		\hline		
    		$P_\mathrm{T}$ & 1 W & $\alpha$ & 4\\
    		$W$ & 10 MHz & $D_\mathrm{BH}$ & 200 ms \\
    		$\rho$ & 50 /km$^2$ & $\lambda$ & 500 /km$^2$\\
    		$\sigma^2$ & -105 dBm/MHz & $I_\mathrm{1}$ & -75 dBm/MHz\\
    		$I_\mathrm{2}$ & -70 dBm/MHz & $I_\mathrm{3}$ & -68 dBm/MHz\\
    		$F$ & 1000 & $\nu$ & 1 \\	
    		${s}_f$ & 1000 & $L$ & 1000 bit\\	
    		%$U_\mathrm{MBH}$ & 20 Mbps & $U_\mathrm{SBH}$ & 1 Mbps \\
    		\hline
    		\hline
    	\end{tabular}
    \end{table}
    
    \begin{figure}[!t]
    	\centering
    	\includegraphics[width=2.5in]{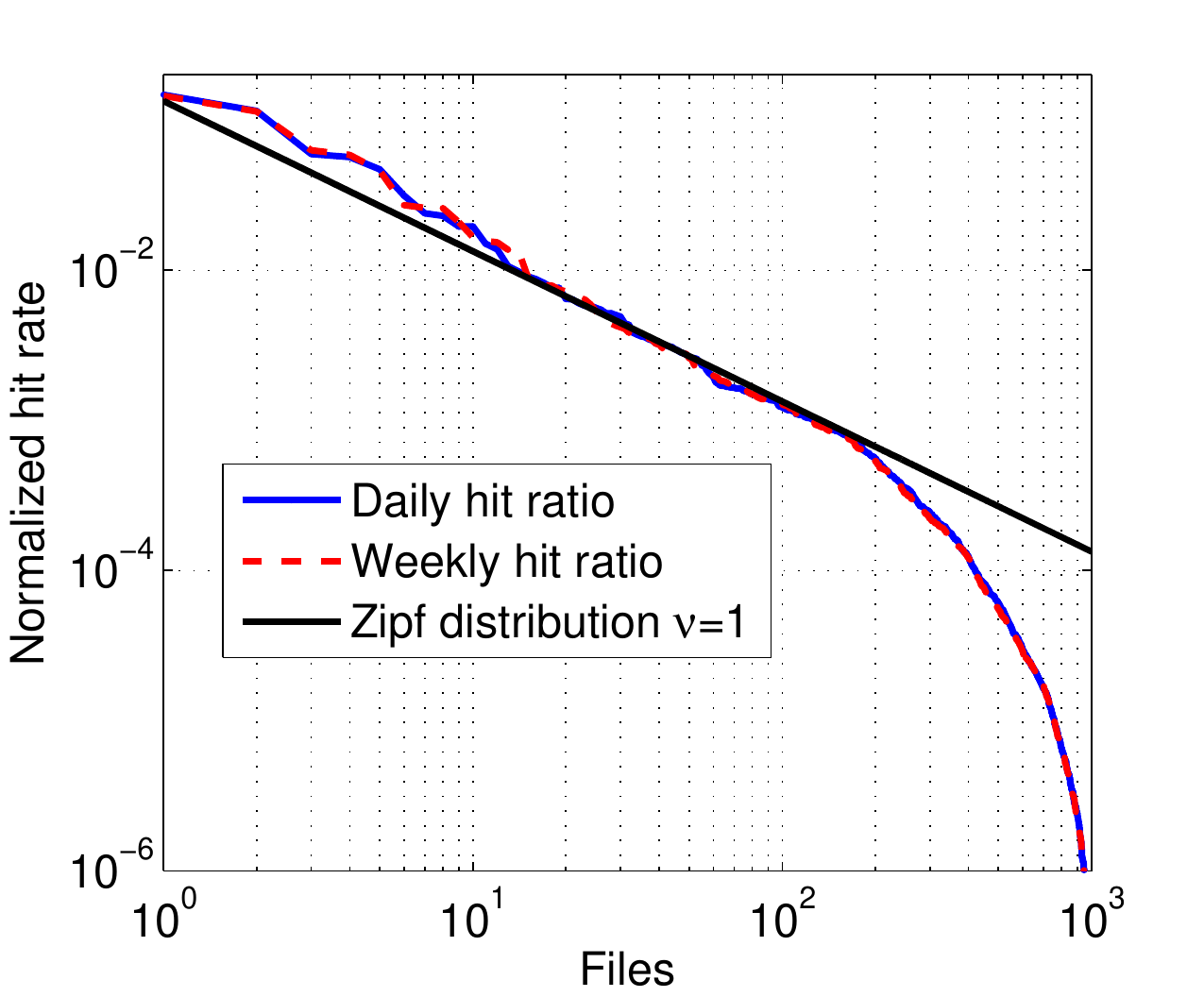}
    	\caption{{YouTube video popularity illustration.}}
    	\label{fig_YouTube_popularity}
    \end{figure}
    
    \begin{figure}[!t]
    	\centering
    	\includegraphics[width=2.5in]{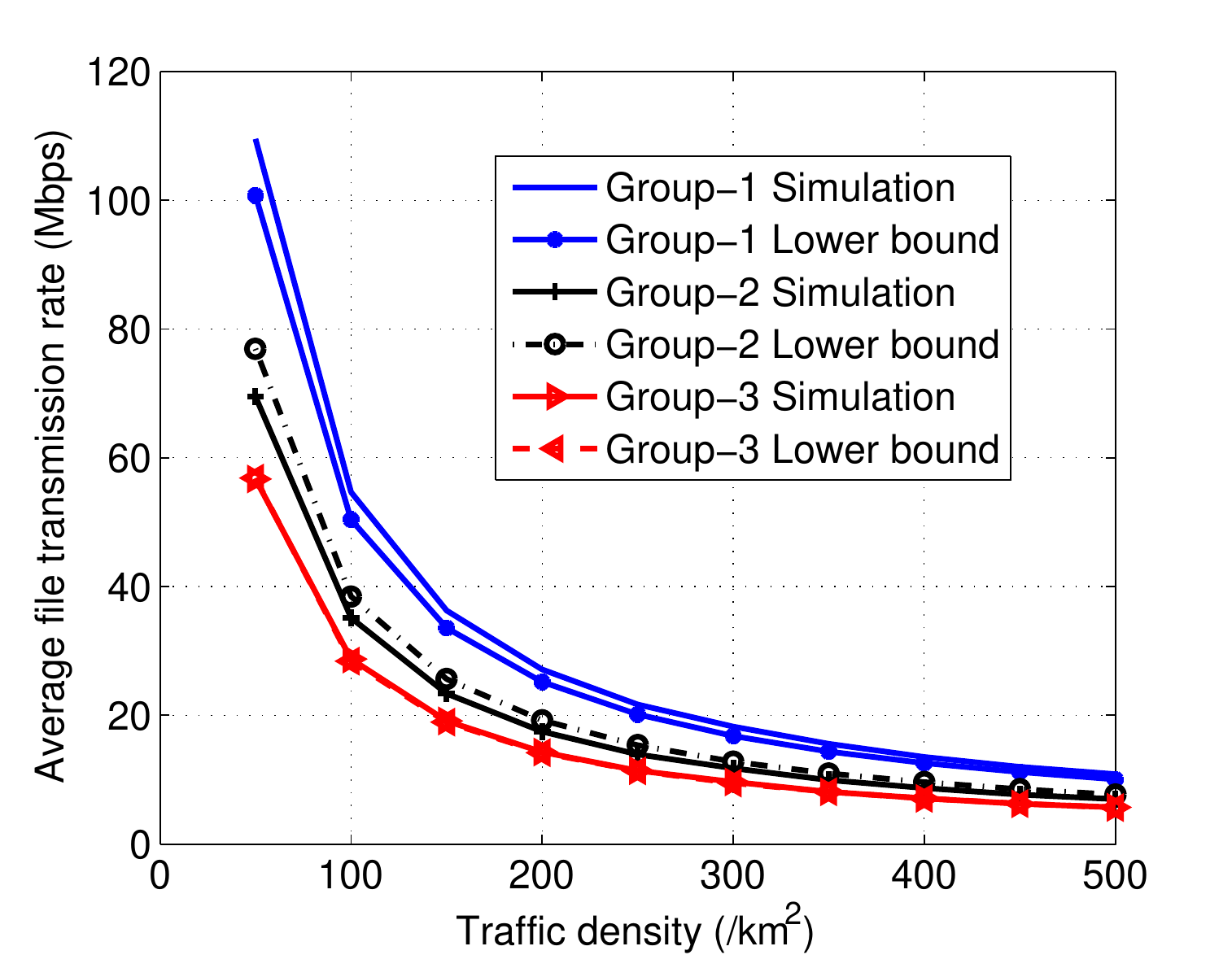}
    	\caption{{Evaluation of average file transmission rate.}} 
    	\label{fig_evaluation}
    \end{figure}
    
    \begin{figure*}[!t]
    	\centering
    	\subfloat[Content hit ratio]{\includegraphics[width=1.8in]{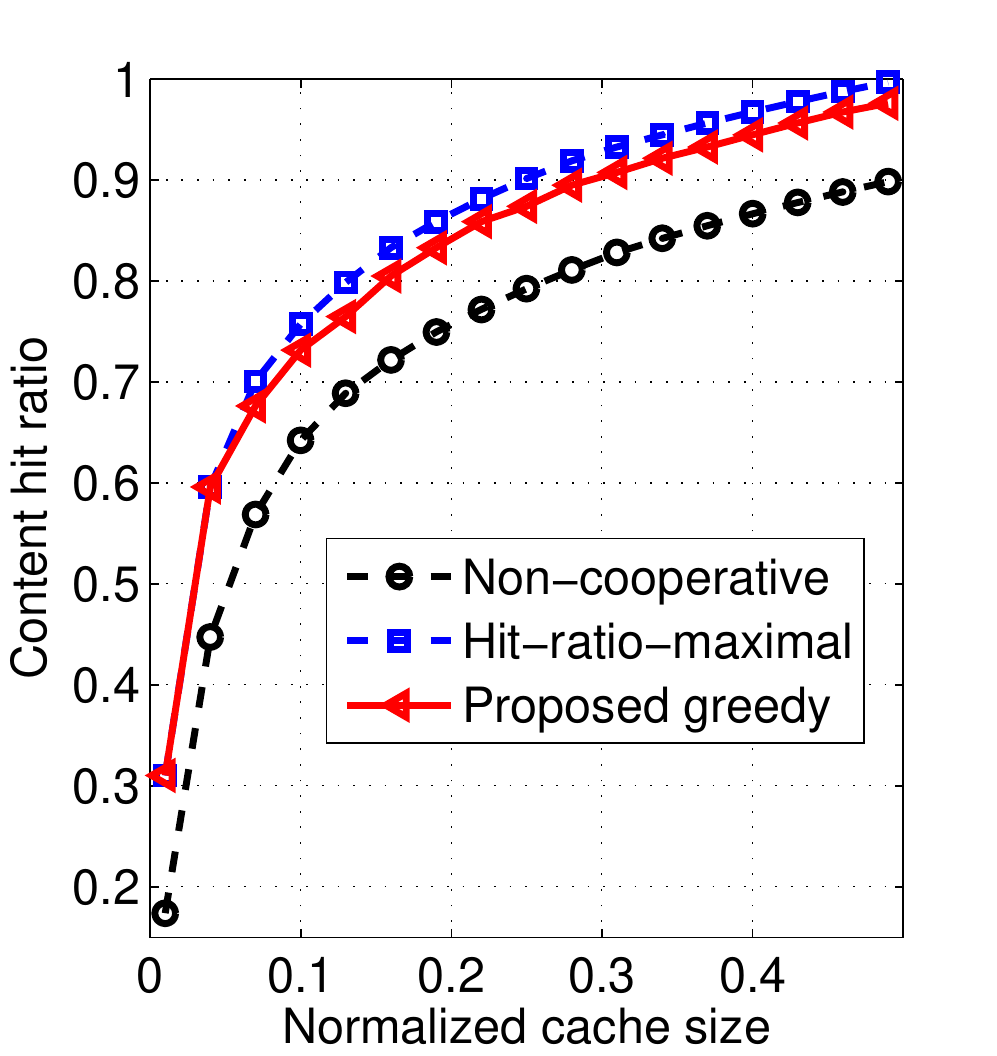}}
    	\hspace{0.1in}
    	\subfloat[Spectrum efficiency]{\includegraphics[width=1.8in]{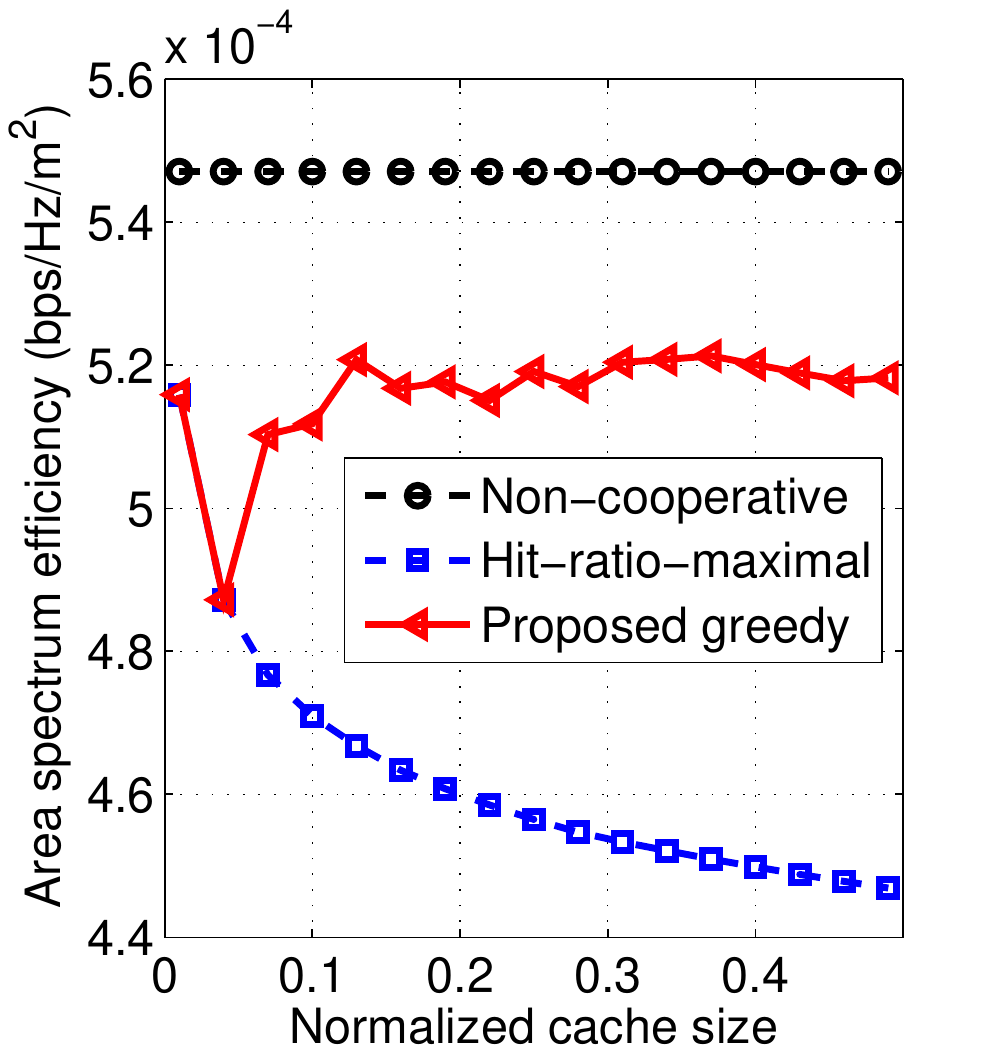}}
    	\hspace{0.1in}
    	\subfloat[Average delay] {\includegraphics[width=1.8in]{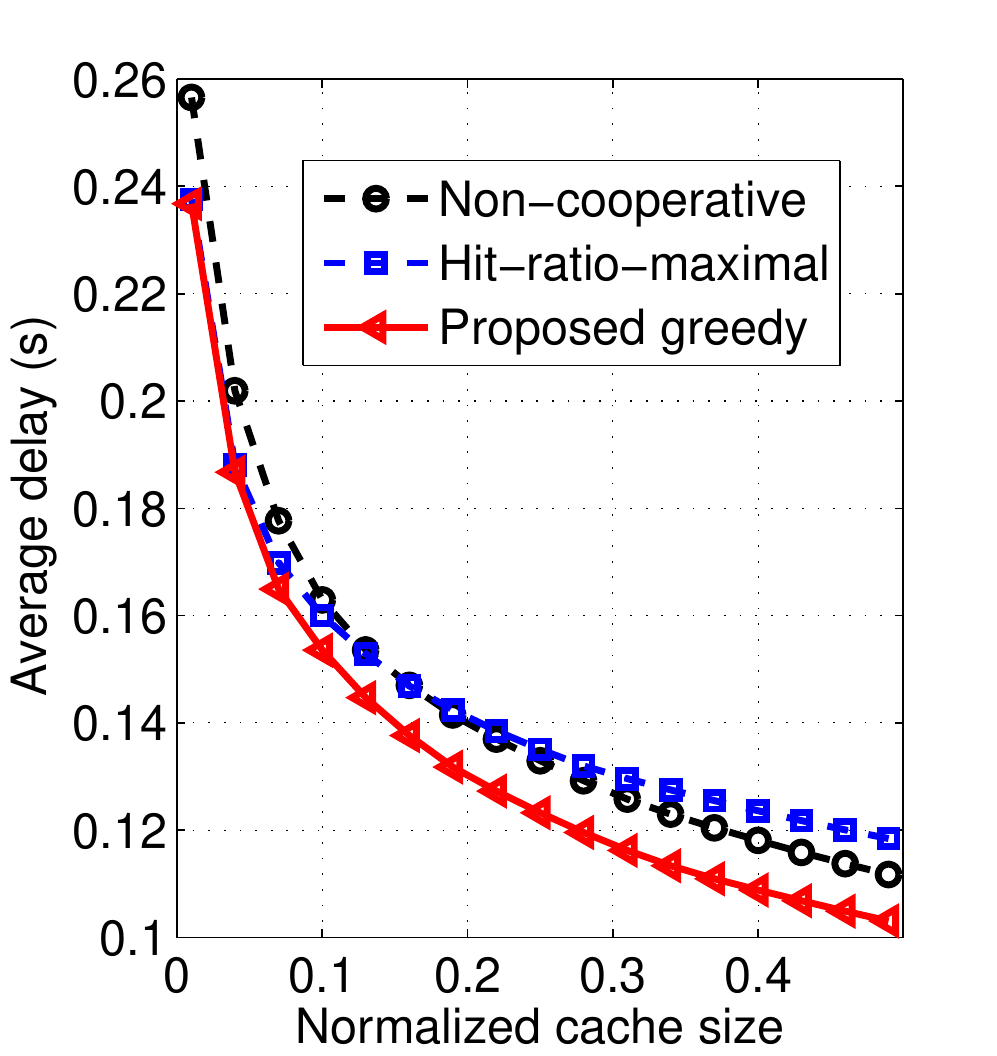}}
    	\caption{{{Performance evaluation of greedy content placement.}}}
    	\label{fig_compare}
    \end{figure*}
    
    In this section, we validate the obtained analytical results of average rates based on extensive system-level simulations, evaluate the performance of the proposed greedy caching algorithm, and study the influence of system parameters as well as cluster size.
    Important simulation parameters are set as Table~\ref{tab_parameter}.
    {{Both real-world data trace and Zipf popularity distribution are adopted. 
    		The real-world data trace is crawled from YouTube where some video owners made their video view statistics open to public, and the view amount information is recorded on a daily basis. 
    		We randomly crawled 1000 videos on May 2017. The most popular video has been watched over 9 million times during the May of 2017, while the least popular one has been rarely viewed.
    		The view amounts (in both daily and weekly scales) of the 1000 videos are normalized to represent the hit ratio, illustrated as Fig.~\ref{fig_YouTube_popularity}.
    		Although the popularity of a specific file may change with time, the overall popularity distribution of the whole file library is shown to be stable in both daily and weekly scales.
    		We adopt the normalized daily content popularity distribution to evaluate the proposed greedy content placement algorithm.
    		The Zipf distribution has been widely adopted to model the content popularity distribution \mbox{\cite{video_popularity_2009}}:}}
    \begin{equation}
    q_f = \frac{1/f^\nu}{\sum_{h=1}^{F} 1/h^\nu},
    \end{equation}
    where $\nu \geq 0$ represents the skewness of popularity distribution, and a larger $\nu$ corresponds to more concentrated file requests.
    Thus, we vary the skewness parameter $\nu$ in simulation to depict diverse applications and contents.
    {{As shown in Fig.~\ref{fig_YouTube_popularity}, the normalized content hit rate of the top 10\% of real trace YouTube videos can be approximated by the Zipf distribution with $\nu=1$.
    		The other files can be outdated ones and rarely requested, causing the mismatch.
    		As these files do not need to be considered for edge caching, the Zipf distribution can be applied to model the popularity distribution of real-world data trace.}}

    \subsection{Analytical Results Evaluation}
    
    {{The derived lower bound of average file transmission rates for radio access is validated, shown as Fig.~\ref{fig_evaluation}, where Group-$k$ in the legend represents the results when users are associated with the $k$th candidate SBS.
    		The results of lower bound is based on Eqn.~(\ref{eq_r_aver_k}).}}
    The simulation results are calculated based on the Monte Carlo method, whereby the SBS topology, user locations, and channel fading are generated randomly according to the corresponding probability distribution functions.
    Fig.~\ref{fig_evaluation} shows that the derived lower bounds of average transmission rate are quite close to the simulation results for different groups of users, and both decrease with the traffic density due to the limited radio resources.
    Therefore, Lemma~1 is validated, and the derived lower bound can be applied to approximate the average file transmission rate for theoretical analysis.
    \begin{figure*}[!t]
    	\centering
    	\subfloat[Content popularity] {\includegraphics[width=2.5in]{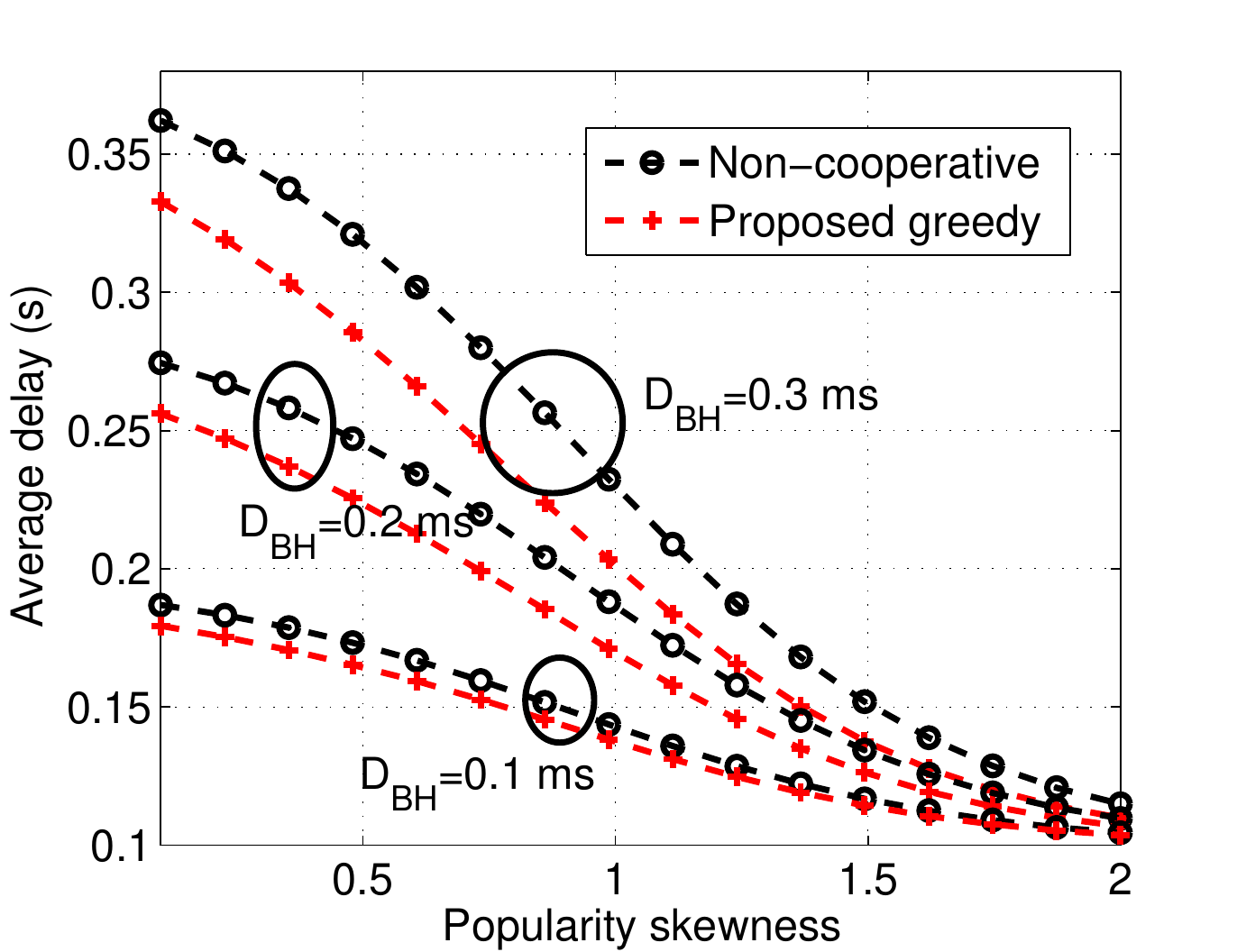}}
    	\hspace{0.5in}
    	\subfloat[Small cell density]{\includegraphics[width=2.5in]{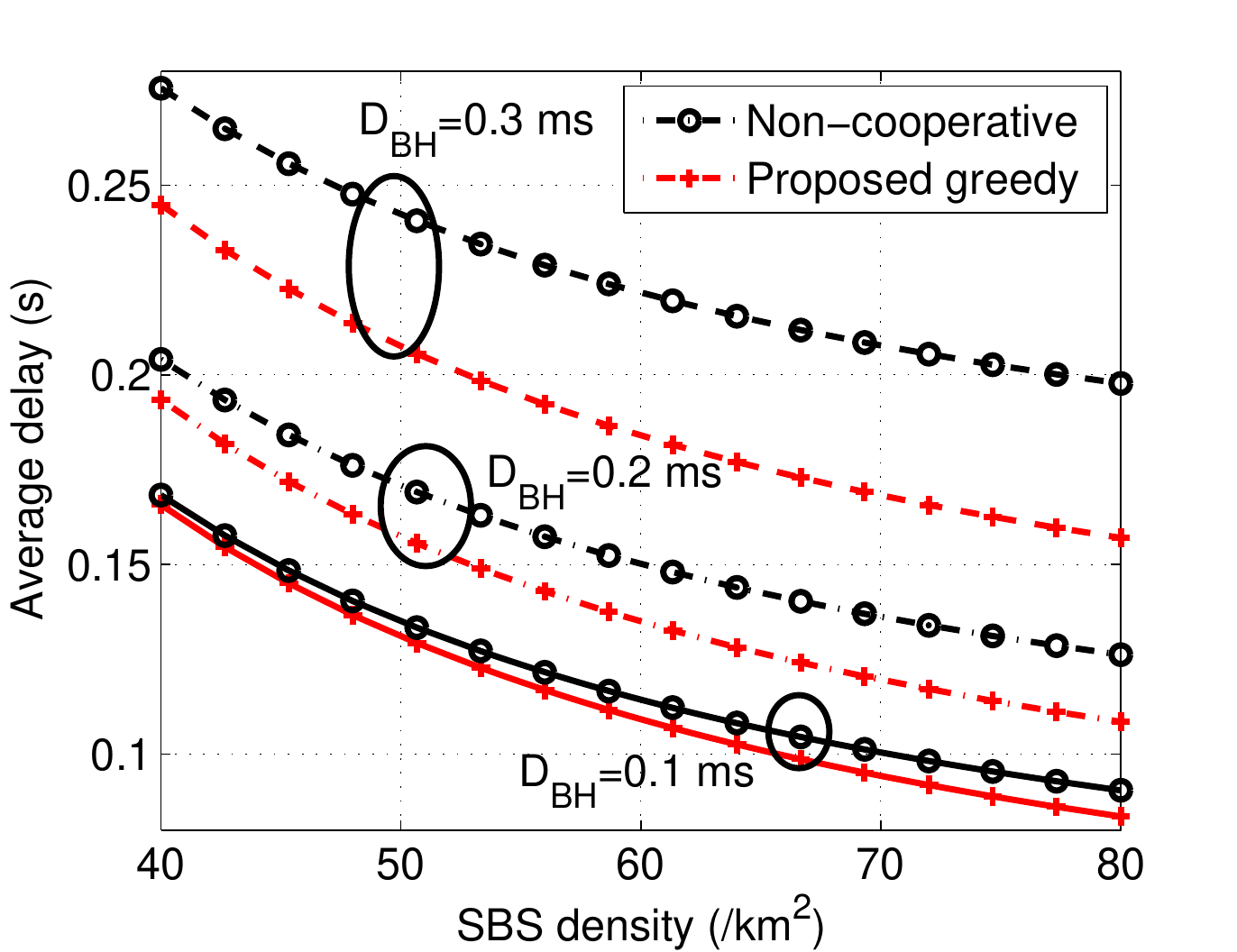}}
    	\caption{{{Influence of system parameters.}}}
    	\label{fig_influence}
    \end{figure*}
    
    \subsection{Tradeoff between Content Diversity and Spectrum Efficiency}
    
    {{Fig.~\ref{fig_compare} demonstrates the tradeoff between content diversity and spectrum efficiency, where different caching schemes are compared in terms of content hit ratio, spectrum efficiency and average delay.
    		The cluster size is set as 2, and the daily YouTube video hit ratio is used.
    		Two caching schemes are adopted as benchmarks.
    		%\footnote{All the three schemes adopt the optimal bandwidth allocation given by Proposition~1.}.
    		Under the non-cooperative scheme, users are only served by the nearest SBS to achieve the high transmission rate, where each SBS cache $s_f$ segments of the $C/{s}_f$ most popular files.
    		On the contrary, the hit-ratio-maximal scheme exploits caching diversity to maximize local content hit ratio, where each SBS cache $s_f/K$ segments of the $C K / s_f$ most popular files.
    		In fact, maximizing the content hit ratio has been considered as a key objective of edge caching, which is favorable in dealing with backhaul congestion and cache size limitations \mbox{\cite{Chen17_CoMP_cluster_cache_TWC, Serbetci17_hit_rate_multi_PPP_WCNC}}.	}}
    
    {{As shown in Figs.~\ref{fig_compare}~(a) and (b), the hit-ratio-maximal scheme always achieves the highest content hit ratio with the lowest spectrum efficiency.
    		Specifically, the content hit ratio increases with the cache size, whereas the spectrum efficiency decreases.
    		The reason is that more users fetch file segments from the cache of farther SBSs as cache size increases, introducing larger wireless transmission delay.
    		On the contrary, the non-cooperative caching scheme is shown to maintain the highest spectrum efficiency but the content hit rate is the lowest, regardless of the cache size.
    		This result is rational since all users are only served by the home SBSs, which can provide maximal average transmission rate.	
    		Compared with the other two schemes, the proposed greedy scheme is shown to balance content hit rate and spectrum efficiency under different cache sizes.	
    		Fig.~\ref{fig_compare} (c) demonstrates that the average file transmission delay decreases with cache size, under all the three schemes.
    		The greedy scheme presents the minimal transmission delay, while the hit-ratio-maximal algorithm performs better than the non-cooperative scheme only when the cache size is smaller than some threshold.}}
    
    {{Figs.~\ref{fig_compare} (a), (b) and (c) are consistent with the analysis.
    		Specifically, the proposed greedy scheme balances content diversity and spectrum efficiency to minimize the average file transmission delay.
    		% by adjusting the content placement and the corresponding traffic load distribution.
    		When the cache size is small, enhancing content diversity is more important due to backhaul congestion, and the proposed greedy scheme is equivalent to the hit-ratio-maximal scheme.
    		When the cache size further increases, the backhaul congestion has been significantly relieved with local caching, and increasing content hit ratio through traffic steering is no longer advantageous considering the degraded spectrum efficiency.
    		In this case, the proposed greedy scheme maintains the spectrum efficiency at certain level, by slightly sacrificing the content hit rate.
    	}}
    	
    	%In general, cooperative caching can increase the content hit rate by steering users to the cache of farther SBSs, which however degrades the spectrum efficiency due to increased path loss.
    	%Thus, a good cooperative caching scheme should balance the content diversity and spectrum efficiency, based on system parameters such as cache size.
    	%The simulation results shown in Fig.~\ref{fig_compare} are consistent with the analytical results.

    	\subsection{Impact of System Parameters}
    	
    	{{To evaluate the performance of the proposed greedy scheme under different traffic demands and use scenarios, we further study the influences of key system parameters, including content popularity distribution, network density, and backhaul delay, illustrated as Fig.~\ref{fig_influence}.
    			The Zipf popularity distribution is adopted in Fig.~\ref{fig_influence}~(a), where popularity skewness varies to represent different applications.
    			Fig.~\ref{fig_influence}~(b) studies the influence of SBS density based on daily trace, corresponding to different network scenarios like dense urban or sparse rural networks.}}
    	The results reveal three facts.
    	Firstly, the performance gain of greedy algorithm decreases with the content popularity skewness, shown as Fig.~\ref{fig_influence}~(a).
    	This is because that the content hit ratio gain brought by cooperative caching degrades when the content requests are more concentrated.
    	Secondly, the performance gain of greedy algorithm increases with cell density, shown as Fig.~\ref{fig_influence}~(b).
    	As the network gets more densified, the cost of traffic steering decreases due to reduced transmission distance and path loss, enhancing the benefit of cooperative caching.
    	Thirdly, the greedy algorithm is even more advantageous when backhaul delay is high, as shown in both Figs.~\ref{fig_influence}~(a) and (b).
    	This result actually indicates the effectiveness of cooperative caching on reducing backhaul transmissions.
    	With the three observed facts, we can conclude that the proposed greedy algorithm can be more effective to cache the less concentrated contents in denser networks with higher backhaul delay.
    	
    	\subsection{Optimal Cluster Size}

    	\begin{figure*}[!t]
    		\centering
    		\subfloat[With respect to content popularity] {\includegraphics[width=2.5in]{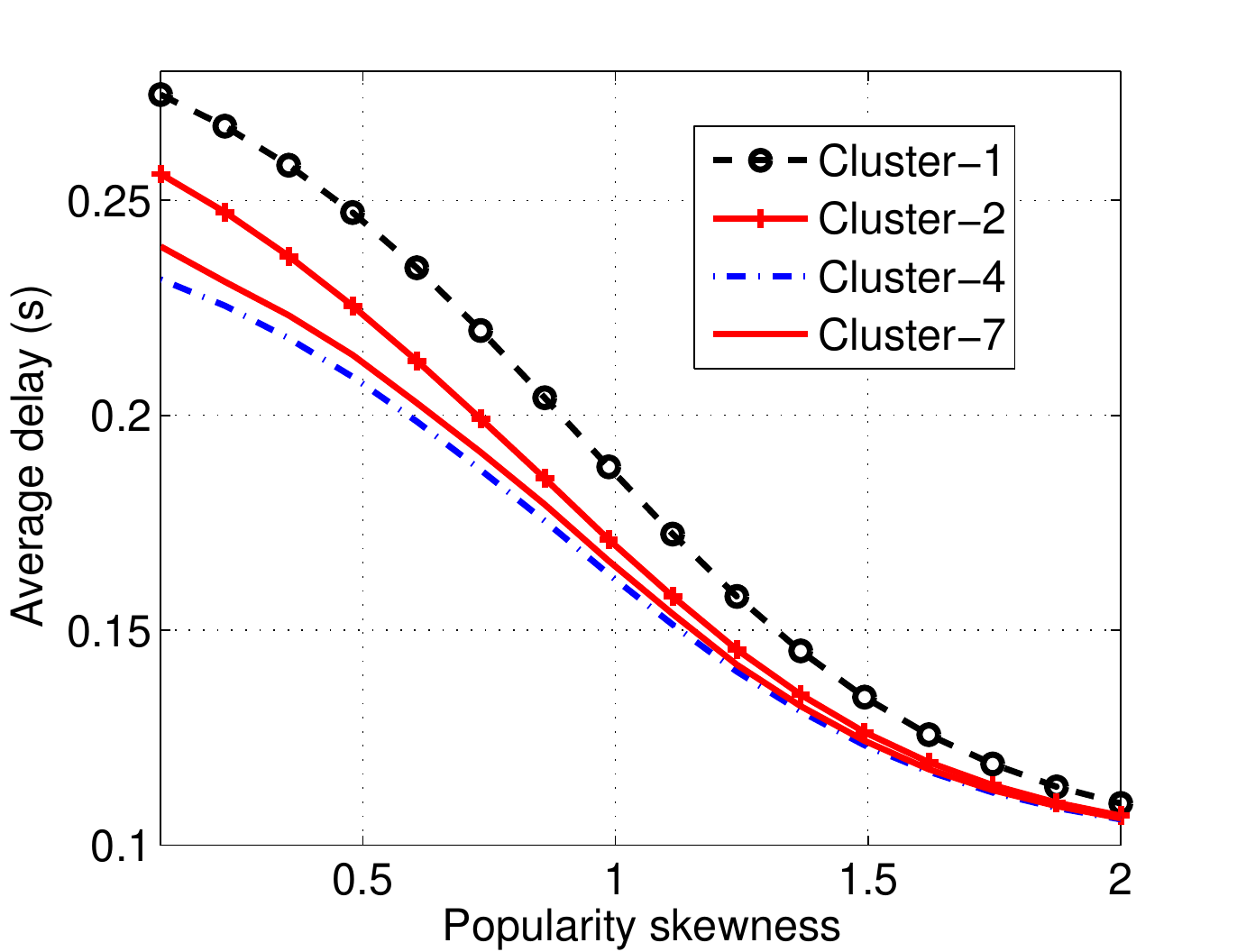}}
    		\hspace{0.5in}
    		\subfloat[With respect to small cell density]{\includegraphics[width=2.5in]{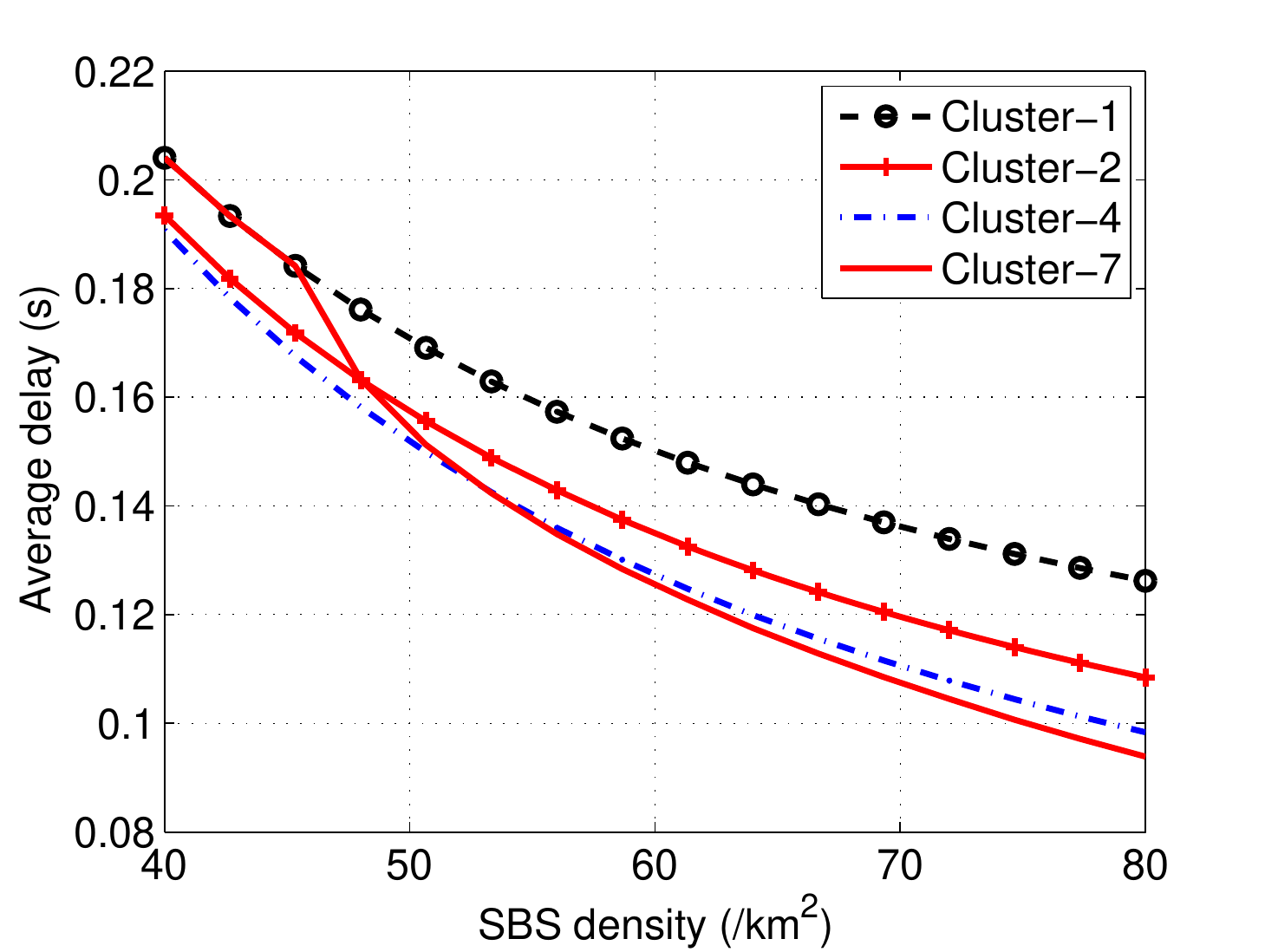}}
    		\caption{{{Performance with different cluster size $K$.}}}
    		\label{fig_influence_cluster_size}
    	\end{figure*}
    	
\begin{figure}[!t]
    		\centering
    		\includegraphics[width=2.5in]{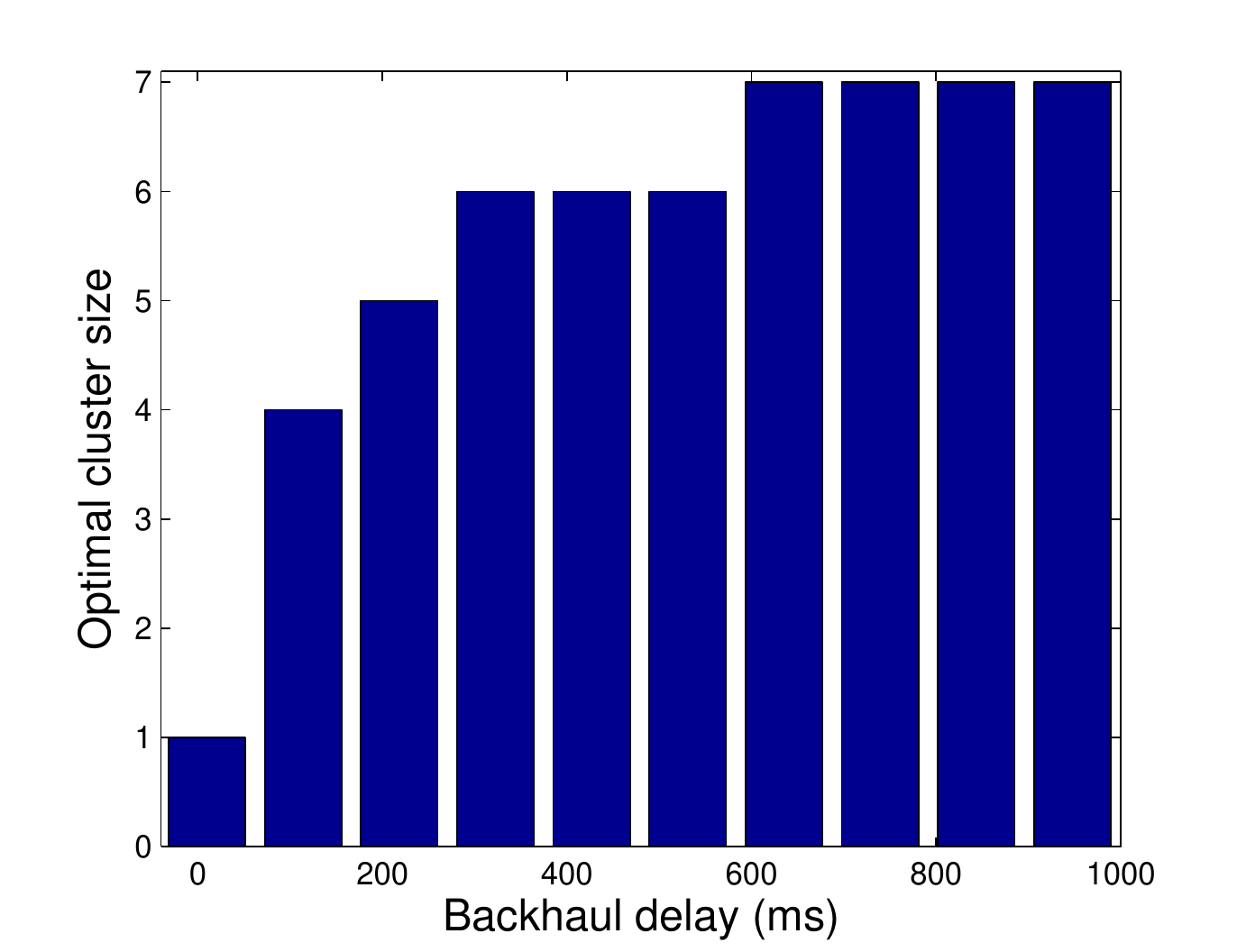}
    		\caption{{{Optimal cluster size with respect to backhaul delay.}}}
    		\label{fig_optimal_clustering}
    	\end{figure}

    	{{As analyzed in previous context, the cluster size also influences the content diversity and spectrum efficiency.}}
    	Fig.~\ref{fig_influence_cluster_size} reveals the system performance with different cluster sizes.
    	As shown in Fig.~\ref{fig_influence_cluster_size}(a), the average delay can be minimized when the cluster size is {{4}}.
    	In addition, when the cluster size exceeds {{7}}, the average delay decreases with the cluster size, indicating that the benefit of cooperative caching will vanish if the cluster size is too large.
    	In this case, it is optimal to store the most popular files as a whole, to avoid users served by SBSs farther away.
    	The simulation results are consistent with the analytical ones.
    	In fact, the cluster size also influences the tradeoff between caching diversity and spectrum efficiency.
    	As cluster size increases, users can fetch files from more SBSs, which increases caching diversity whereas degrades spectrum efficiency.
    	Accordingly, the optimal cluster size should balance the tradeoff to minimize average transmission delay.
    	The optimal cluster size also depends on the SBS density, shown as Fig.~\ref{fig_influence_cluster_size}(b).
    	Specifically, larger cluster sizes are more advantageous in denser networks, due to the reduced transmission distance and path loss on average.
    	{{For example, the optimal cluster size is shown to be 4 when the SBS density is smaller than 54 /km$^2$, but becomes 7 when the SBS density further increases.}}
    	Consider another example when cluster size equals to 7.
    	The average delay is shown to be no smaller than that of the non-cooperative caching when the SBS density is smaller than {{45 /km$^2$.}}
    	In this case, the condition of Proposition~2 (i.e., Eqn.~(\ref{eq_proposition_2})) cannot be satisfied, as the caching diversity gain is overwhelmed by the spectrum efficiency degradation due to the long transmission distance.
    	However, the average delay decreases significantly when the SBS diversity exceeds {{45 /km$^2$.}}
    	Then, the condition of Proposition~2 holds with reduced spectrum efficiency degradation, and thus cooperative caching improves delay performance.
    	Notice that the optimal cluster size also increases with the backhaul delay, shown as Fig.~\ref{fig_optimal_clustering}.
    	{{The reason is that increasing content hit ratio is more advantageous when backhaul delay is higher, which can be realized by increasing cluster size. }}
    	
    	\begin{figure}[!t]
    		\centering
    		\includegraphics[width=2.5in]{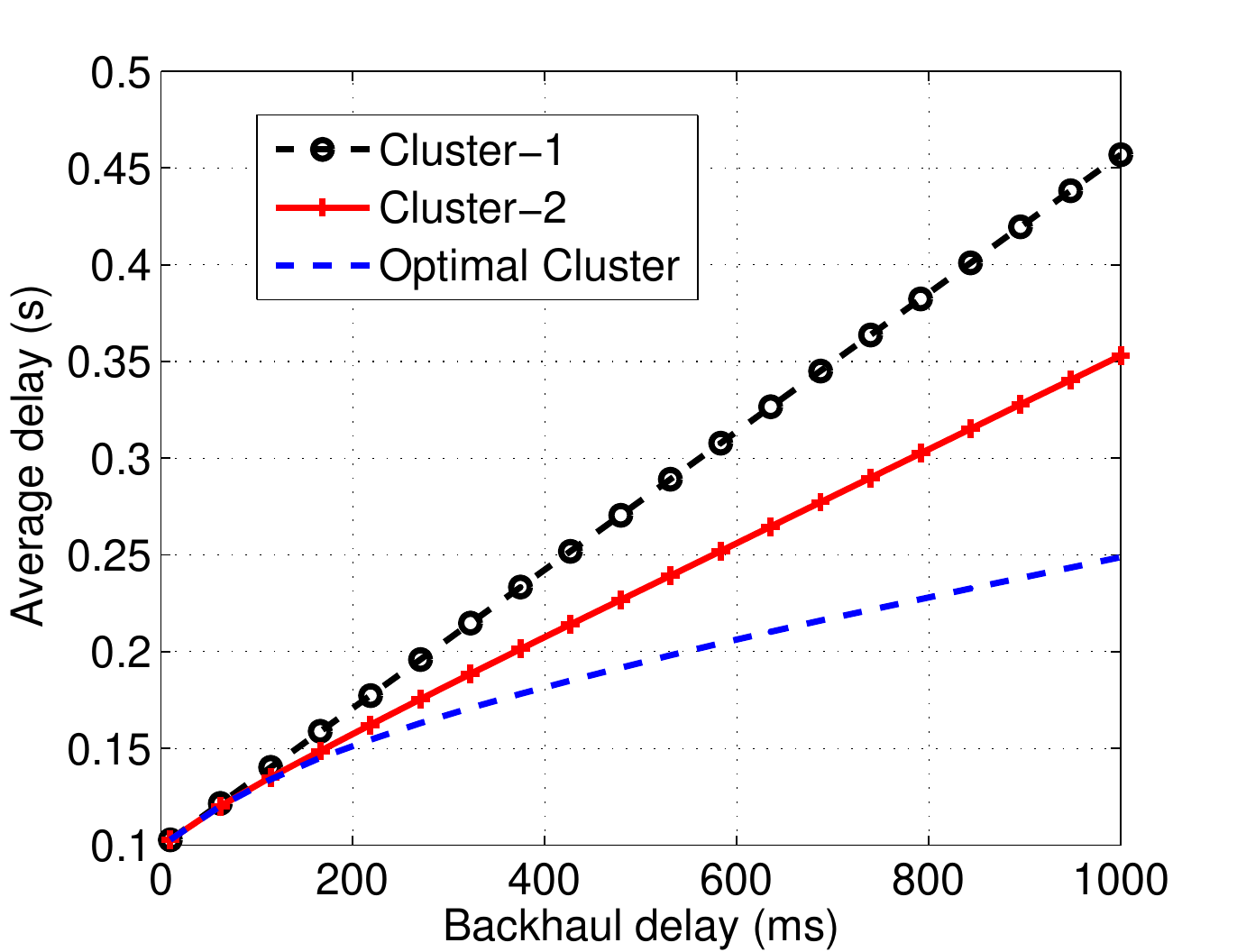}
    		\caption{{{Delay enhancement with optimal clustering (cluster size denoted as $K$).}}}
    		\label{fig_optimal_cluster_gain}
    	\end{figure} 
    	
    	Fig.~\ref{fig_optimal_cluster_gain} further demonstrates the effectiveness of optimizing cluster size.
    	{{The average transmission delay increases with the backhaul delay regardless of cluster size, whereas the increasing rates are different.
    			When the cluster size is a constant, the average transmission delay is shown to increase almost linearly with backhaul delay.
    			When the cluster size is optimized, the average transmission delay increases sub-linearly.
    			For illustration, compared with the non-cooperative caching (cluster size set to 1), the average delay can be reduced by around {{25\% and 45\%}} through cluster size optimization, when the backhaul delay is 400 ms and 1 s, respectively.
    			The important insights for application is that the cluster size should be adjusted based on the system parameter and status.
    			For instance, when the backhaul is congested during rush hours, the user-centric cluster size should enlarge to increase content hit ratio and reduce backhaul pressure.
    			Instead, when the traffic load decreases at midnight, users can just fetch files from home SBSs with the cluster size shrink to 1.}}
    	
%%%%%%%%%%%%%%%%%%%%%%%%%%%%%%%%%%%%%%%%%%%%%%%%%%%%%%%%%%%%%%%%%%%%%%%%%%%%%%%%%%%%%%%%%%%%%%%%%%%
\section{Conclusions and Future Work}
    \label{sec_conclusion}
	
	In this paper, the cooperative edge caching has been investigated in large-scale user-centric clustered mobile networks, aiming at minimizing the average file transmission rates with caching size and bandwidth constraints.
	Based on the optimal bandwidth allocation obtained by the Lagrange multiplier method, a linear-complexity greedy content placement algorithm has been proposed with guaranteed performance. 
	In addition, an explicit condition constraining the maximal cluster size has been obtained, which offers a guideline for user-centric clustering in practical networks.
	The results of the optimal content placement and SBS clustering have both revealed the tradeoff relationship between content diversity and spectrum efficiency with cooperative edge caching.
	For our future work, we will investigate the impact of user mobility and unknown content popularity on mobile edge caching.
	{{Furthermore, it is also interesting to design advanced cooperative caching scheme with instant information of cell load and channel condition in heterogeneous networks with diversified backhaul capacity.}}

%%%%%%%%%%%%%%%%%%%%%%%%%%%%%%%%%%%%%%%%%%%%%%%%%%%%%%%%%%%%%%%%%%%%%%%%%%%%%%%%%%%%%%%%%%%%%%%%%%%

%\appendices
%   \input{appendix.tex}
   
%   % use section* for acknowledgment
%   \ifCLASSOPTIONcompsoc
%   % The Computer Society usually uses the plural form
%   \section*{Acknowledgments}
%   \else
%   % regular IEEE prefers the singular form
%   \fi

\ifCLASSOPTIONcaptionsoff
   \newpage
   \fi

%
%
%\vspace{-20mm}

\end{document}